\documentclass[a4paper,twoside]{article}

\newcommand{\affil}[1]{$^{\rm #1}$}

\usepackage{graphicx,amssymb}
\usepackage{epsf,psfig}

\date{}

\title{\large\bf\flushleft Trapped and escaping orbits in an axially symmetric galactic type potential}

\author{\parbox{\textwidth}{\flushleft
\vspace{-0.5 cm}
{\it Euaggelos E. Zotos\affil{}}\\
\vspace{0.4 cm}
{\small \affil{}\,Department of Physics, \\\
Section of Astrophysics, Astronomy and Mechanics, \\\
Aristotle University of Thessaloniki, \\\
GR-541 24, Thessaloniki, Greece}\\
{\small \affil{}\,Email: evzotos@astro.auth.gr}}}

\begin{document}

\twocolumn[
\begin{changemargin}{.8cm}{.5cm}
\begin{minipage}{.9\textwidth}
\vspace{-1cm}
\maketitle

\small{\bf Abstract:}

In the present article, we investigate the behavior of orbits in a time independent axially symmetric galactic type potential. This dynamical model can be considered to describe the motion in the central parts of a galaxy, for values of energies larger than the energy of escape. We use the classical method of the surface of section, in order to visualize and interpret the structure of the phase space of the dynamical system. Moreover, the Lyapunov Characteristic Exponent (L.C.E), is used in order to make an estimation of the degree of the chaoticity of the orbits in our galactic model. Our numerical calculations suggest that in this galactic type potential, there are two kinds of orbits: (i) escaping orbits and (ii) trapped orbits which do not escape at all. Furthermore, a large number of orbits of the dynamical system, display chaotic motion. Among the chaotic orbits, there are orbits that escape fast and also orbits that remain trapped for vast time intervals. When the value of the test particle's energy exceeds slightly the energy of escape, the amount of the trapped regular orbits increases, as the the value of the angular momentum increases. Therefore, the extent of the chaotic regions observed in the phase plane decreases as the value of the energy increases. Moreover, we calculate the average value of the escape period of the chaotic orbits and we try to correlate it with the value of the energy and also with the maximum value of the $z$ component of the orbits. In addition, we find that the value of the L.C.E corresponding to each chaotic region, for different values of the energy, increases exponentially as the value of the energy increases. Some theoretical arguments in order to support the numerically obtained outcomes are presented.


\medskip
\medskip
\end{minipage}
\end{changemargin}
]
\small

\section{Introduction}

The shape of galaxies strongly depends on their orbital populations. Stars within galaxies belong to orbital families. In particular, the size and the shape of a galaxy determines the relative populations of these families. As a galaxy evolves, capture and escape of stars between these families takes place. Therefore, capture and escape are generic processes that will have occurred many times in the lifetime of galaxies.

Each orbital family has a parent periodic orbit - that is, an orbit that describes a closed figure. For example, in a spherical galaxy, all stars belong to the family of tube orbits, whose members librate around the closed circular orbits. An oval distortion in the center of the galaxy is supported by the family of box orbits, whose parent periodic orbits are the radial orbits. As the distortion grows, stars are transferred from the loop to the box orbital family. This capture process is important in the formation, maintenance and secular evolution of non-axisymmetric structure, such as galactic bars, rings and spiral arms (e.g., Kalnajs 1973; Contopoulos, H\'{e}non \& Lynden-Bell 1973; Tremaine \& Weinberg 1984). Whether or not a trapped star remains trapped, may depend on the presence of a central massive black hole, or a simple mass concentration. Close encounters can scatter stars away from their initial orbits and thus cause a gradual disruption of the population of the orbits.

The phenomenon of escaping stars from stellar systems, has been an active field of research during the last decades (e.g., Contopoulos 1990; Contopoulos \& Kaufmann 1992; Siopis et al. 1995a, 1995b; Kandrup et al. 1999; Fukushige \& Heggie 2000; Contopoulos \& Efstathiou 2004; Contopoulos \& Harsoula 2005; Contopoulos \& Patsis 2006; Papadopoulos \& Caranicolas 2007). The reader can find more details on the subject of escapes in the review of Siopis et al. 1996 and also in the book of Contopoulos 2002. When a star has a value of energy higher than the energy of escape, the equipotential curves or, what is equivalent, the zero velocity curves (ZVCs) are open and there are cases, where a star can escape from the stellar system. On the other hand, there are also cases, where stars which hold values of energy much more larger than the energy of escape do not escape at all, even though the ZVCs are open and therefore at least one channel of escape do exists. A characteristic example of such non escaping stars, are those moving in orbits with initial conditions close to those of a stable periodic orbit. Apart from galactic dynamics, the phenomenon of escaping and trapped orbits, has been also studied in the case of the three body problem (e.g., Anosova 1986; Benet et al. 1996, 1998), as well as in mappings (Bleher et al. 1988).

The main scope of the present research, is to shed light on the properties of motion in a dynamical system with trapped and escaping orbits. Especially, we shall study the nature of motion (regular or chaotic) for values of energy larger than the energy of escape, that is the case when the ZVC is open. In this case, there are regular orbits which do not escape at all. We shall give the set of initial conditions leading to this kind of orbits as a function of the energy. What is much more interesting, it that there are two different kinds of chaotic orbits. The first kind of chaotic orbits contains chaotic orbits, which remain trapped inside the ZVC for long time intervals, before escaping to infinity. We shall call these orbits trapped chaotic orbits and investigate their properties, such as their time scale of escape and their degree of chaos. The second kind of chaotic orbits contains chaotic orbits which escapes from the ZVC in very short time intervals. These orbits are called fast escaping orbits. The main feature of these orbits, is that their initial conditions are very close to the neighborhood of the escape channel of the ZVC.

The layout of this article is organized as follows: In Section 2, we present a mathematical analysis of our galactic type model and also a numerically obtained relationship connecting the angular momentum $L_z$ and the energy of escape $h_{esc}$. Moreover, in the same Section we study the areas of initial conditions corresponding to trapped regular orbits, as a function of the energy. In Section 3, we study the structure of the dynamical system, using the outcomes from the phase planes. A numerical relationship between the extent of the chaotic regions and the value of the energy is also presented. In Section 4 we conduct a detailed study on the trapped and escaping chaotic orbits and we also connect the degree of chaos with the value of the energy. We close with Section 5, where a discussion and the conclusions of this research are given and a comparison with previous work is also made.

\section{Description of the dynamical model}

In the present paper, we shall study the properties of motion in the galactic type potential
\begin{equation}
V(r,z) = \frac{\omega ^2}{2}\left(r^2 + z^2 \right) - \epsilon \left[ \alpha \left(r^4 + z^4 \right) + 2 \beta r^2 z^2 \right].
\end{equation}
Potentials of the form of Equation (1), that is potentials made up of perturbed harmonic oscillators, are among the most well studied in non-linear dynamics (see Deprit \& Elipe 1991; Elipe \& Deprit 1999; Elipe 2001). Moreover, this potential has been applied several times in previous research works (e.g., Caranicolas \& Vozikis 1999; Caranicolas 2001; Karanis \& Caranicolas 2002; Caranicolas \& Vozikis 2002; Caranicolas \& Papadopoulos 2003b) and can be derived by expanding global galactic potentials near the central stable equilibrium point of the system, that is the center of the galaxy. In the present study it can be considered to describe the local motion in the meridian $(r,z)$ plane near the central parts of an axially symmetric galaxy. In equation (1) $(r,z)$ are the usual cylindrical coordinates, while $\omega, \epsilon, \alpha$ and $\beta$ are parameters. Such galactic type potentials arise naturally if (a) the density distribution near the galactic center is an analytic function of the coordinates and (b) the Taylor series for the corresponding potential is truncated at fourth order. Here we must point out, that our gravitational galactic type potential is truncated at $r_{max}=1.5$, otherwise the mass density increases outwards from the center, which is practically never observed in galaxies or other stellar systems. Therefore, we study the phenomenon where stars escape from the central parts of a galaxy.

It is well known that the motion of stars in the central parts of almost all galaxies could be satisfactorily represented by the motion of a harmonic oscillator. Therefore, our dynamical model of the perturbed harmonic oscillator potential given by Equation (1) is able to describe local motion of stars in the central region of different kinds of galaxies (i.e. from a non rotating elliptical galaxy to a disk or a spiral galaxy). Once more, we have to point out that the present model can only describe local motion at small distances from the galactic center ($r_{max} \leq 1.5$ kpc).

As the potential $V(r,z)$ is axially symmetric and the $L_z$ component of the angular momentum is conserved, the dynamical structure of the galactic system, can be investigated using the effective potential
\begin{equation}
V_{eff}(r,z) = \frac{L_z^2}{2r^2} + V(r,z),
\end{equation}
in order to study the properties of motion in the meridian $(r,z)$ plane. The equations of motion are
\begin{eqnarray}
\dot{r} = p_r, \ \ \ \dot{z} = p_z, \nonumber\\
\dot{p_r} = - \frac{\partial V_{eff}}{\partial r}, \ \ \ \dot{p_z} = - \frac{\partial V_{eff}}{\partial z},
\end{eqnarray}
where the dot indicates derivative with respect to the time.

The Hamiltonian corresponding to potential (2) is written as
\begin{equation}
H = \frac{1}{2}\left(p_r^2 + p_z^2 \right) + V_{eff}(r,z) = h,
\end{equation}
where $p_r$ and $p_z$ are the momenta per unit mass conjugate to $r$ and $z$ respectively, while $h$ is the numerical value of the Hamiltonian, which is conserved. Equation (4) is an integral of motion, which indicates that the total energy of the test particle is conserved. The Hamiltonian (4), also describes the motion in the $(r,z)$ meridian plane, rotating at the angular velocity
\begin{equation}
\dot{\phi} = \frac{L_z}{r^2}.
\end{equation}

The bulk of the outcomes of the present research are based on the numerical integration of the equations of motion (3). We use a Bulirsh-St\"{o}er integration routine in Fortran 95, with double precision in all subroutines. The accuracy of our results was checked by the constancy of the value of the energy integral (4), which was conserved up to the fifteenth significant decimal figure.

In this work, we use a system of galactic units, where the unit of length is 1 kpc, the unit of time is $10^7$ yr. The velocity unit is 1 kpc/$10^7$ yr = 97.8 km/s and the energy unit is 1 kpc$^2/(10^7$ yr)$^2$. In these units, we take the values: $\omega = 1/(10^7$ yr), $\epsilon = 1/(10^7$ yr kpc)$^2$, $\alpha = 0.2$ and $\beta = -1.2$. The above numerical values of the quantities of the dynamical system are kept constant, during this research and they secure positive density everywhere and free of singularities.

In the case when $L_z=0$, the energy of escape for the potential (1) can be found theoretically (see Caranicolas \& Vozikis 1999 and references therein) and it is equal to
\begin{equation}
h_{esc} = \frac{1}{16 \epsilon \alpha}.
\end{equation}
When $\alpha = 0.2, \beta = -1.2$ and $\epsilon = 1$, we find $h_{esc} = 0.3125$. In the general case where $L_z \neq 0$, the value of the energy of escape can be found numerically. Figure 1 shows a plot of the relation between the angular momentum $L_z$ and the energy $h$. On the upper left part of the $\left[h, L_z\right]$ plane, including the line, the ZVC is closed and the motion of stars is bounded, while on the lower shaded right part of the same diagram the ZVC is open and therefore stars can escape through the open channel. The relation between $h_{esc}$ and $L_z$ is given approximately by the fourth order polynomial equation
\begin{eqnarray}
L_z &=& -55738.3 h_{esc}^4 + 78897.2 h_{esc}^3 - 41862.2 h_{esc}^2 \nonumber \\
&+& 9871.87 h_{esc} - 872.985.
\end{eqnarray}
\begin{figure}[!tH]
\centering
\resizebox{\hsize}{!}{\rotatebox{0}{\includegraphics*{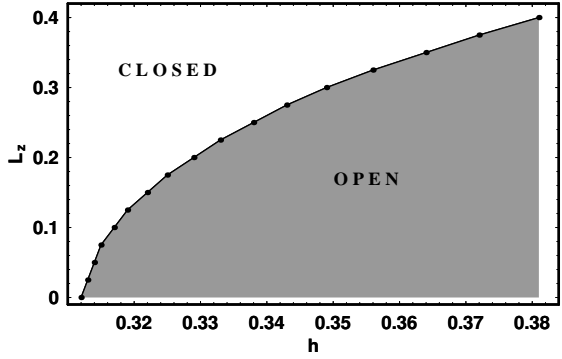}}}
\caption{A plot of the correlation between $L_z$ and $h$. On the upper left part of the $\left[h, L_z\right]$ plane including the solid line the ZVC is closed, while on the lower shaded right part the ZVC is open.}
\end{figure}
\begin{figure}[!tH]
\centering
\resizebox{\hsize}{!}{\rotatebox{0}{\includegraphics*{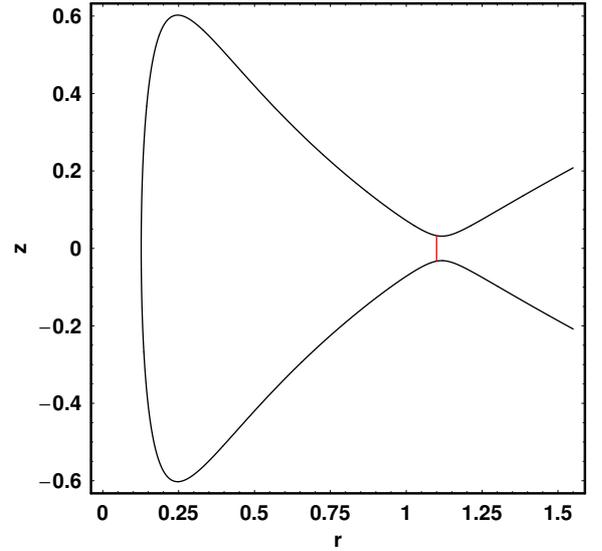}}}
\caption{The ZVC when $h=0.32$ and $L_z=0.10$. The red line represents the Lyapunov orbit.}
\end{figure}

Figure 2 shows the ZVC when $h=0.32$ and $L_z=0.10$. As one can see, there is a channel in the ZVC through which orbits can escape to infinity. The red line represents the Lyapunov orbit. Orbits that cross the Lyapunov orbit with velocity outwards will escape from the dynamical system (Churchill et al. 1979).
\begin{figure}[!tH]
\centering
\resizebox{\hsize}{!}{\rotatebox{0}{\includegraphics*{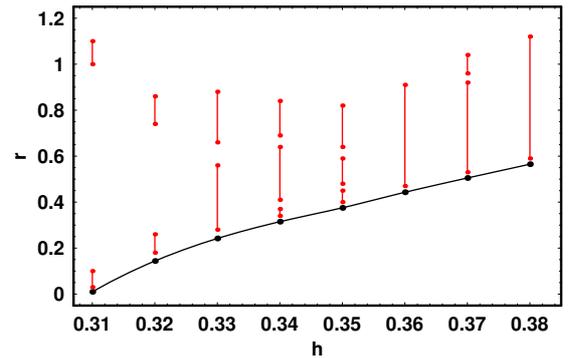}}}
\caption{The black solid line depicts the relationship between $r_{min}$ and $h$, while the red vertical lines correspond to intervals of initial conditions on the $r$ axis of the phase plane corresponding to trapped regular orbits. See text for details.}
\end{figure}

In addition to the escaping orbits, there are regular orbits that do not escape at all. The red vertical lines in Figure 3 show intervals corresponding to initial conditions on the $r$ axis of the $\left(r,p_r\right)$ phase planes $(z=0)$, giving such trapped regular orbits, for each particular value of the energy $h$. We select the pairs of $\left[h, L_{zc}\right]$ in the following way: We take a pair of values from equation (7), say $\left[h, L_{zc}\right] = [0.32, 0.11]$. Then we choose the pair $h = h_{esc}, L_z = L_{zc} - \Delta L$ with $\Delta L = 0.01$. For this pair of values the ZVC is always open. The solid line in Fig. 3 gives the values of $r_{min}$, that is the intersection of the ZVC with the $z=0$ axis. This can be found if we solve numerically the equation
\begin{equation}
V_{eff}(r) = V(r) + \frac{L_z^2}{2r^2} = \frac{1}{2}r^2 - \alpha r^4 + \frac{L_z^2}{2r^2} = h.
\end{equation}
We find that the value of $r_{min}$ increases as the value of the energy $h$ increases. This means that as the energy increases orbits are going far from the center. In fact, what is true is that as $h_{esc}$ and $L_z$ are connected through relation (7), the orbits with larger $L_z$ are not able to approach near the center. It is clear from the diagram of Fig. 3 that the area of the initial conditions on the $r$ axis of the $\left(r,p_r\right)$ phase planes, that is initial conditions $\left(r=r_0, z=z_0=0\right)$, giving regular trapped orbits generally increases as the value of the energy $h$ increases. We shall explain this point in details, later in the next Section, when we will try to connect the extent of the chaotic regions of the phase planes with the value of the energy $h$.

\section{Structure of the dynamical system}

In this Section, we shall use a qualitative but very effective method to reveal the dynamical structure of a Hamiltonian system. In this method, we plot the successive intersections of the orbits using the $\left(r,p_r\right)$, $z=0, p_z>0$ surface of section, in order to determine the regular or chaotic nature of the motion. This particular method, has been extensively applied to Hamiltonian systems with two degrees of freedom, as in these systems the phase plane is a two dimensional plane and therefore, can easily be visualized and interpreted. Here we must note, that this phase plane is not a Poincar\'{e} surface of section, because some orbits have loops that may or may not intersect this surface. A Poincar\'{e} surface of section exists only if orbits intersect a surface like $z=0$ at least once within a certain time interval. However, escaping orbits in general, do not intersect the $z=0$ axis after a certain time, thus the sections of the present paper are not Poincar\'{e} sections. If we set $z=p_z=0$ in equation (4), we obtain the limiting curve in the $\left(r,p_r\right)$ phase plane, which is the curve containing all the invariant curves, for a given value of the energy integral $h$. Thus we choose the values of $r$ and $p_r$ inside the limiting curve, while the value of $p_z$ if found always from the energy integral (4). The limiting curve of the dynamical system corresponds to
\begin{equation}
\frac{1}{2}p_r^2 + V_{eff}(r) = h.
\end{equation}
\begin{figure*}[!tH]
\centering
\resizebox{\hsize}{!}{\rotatebox{270}{\includegraphics*{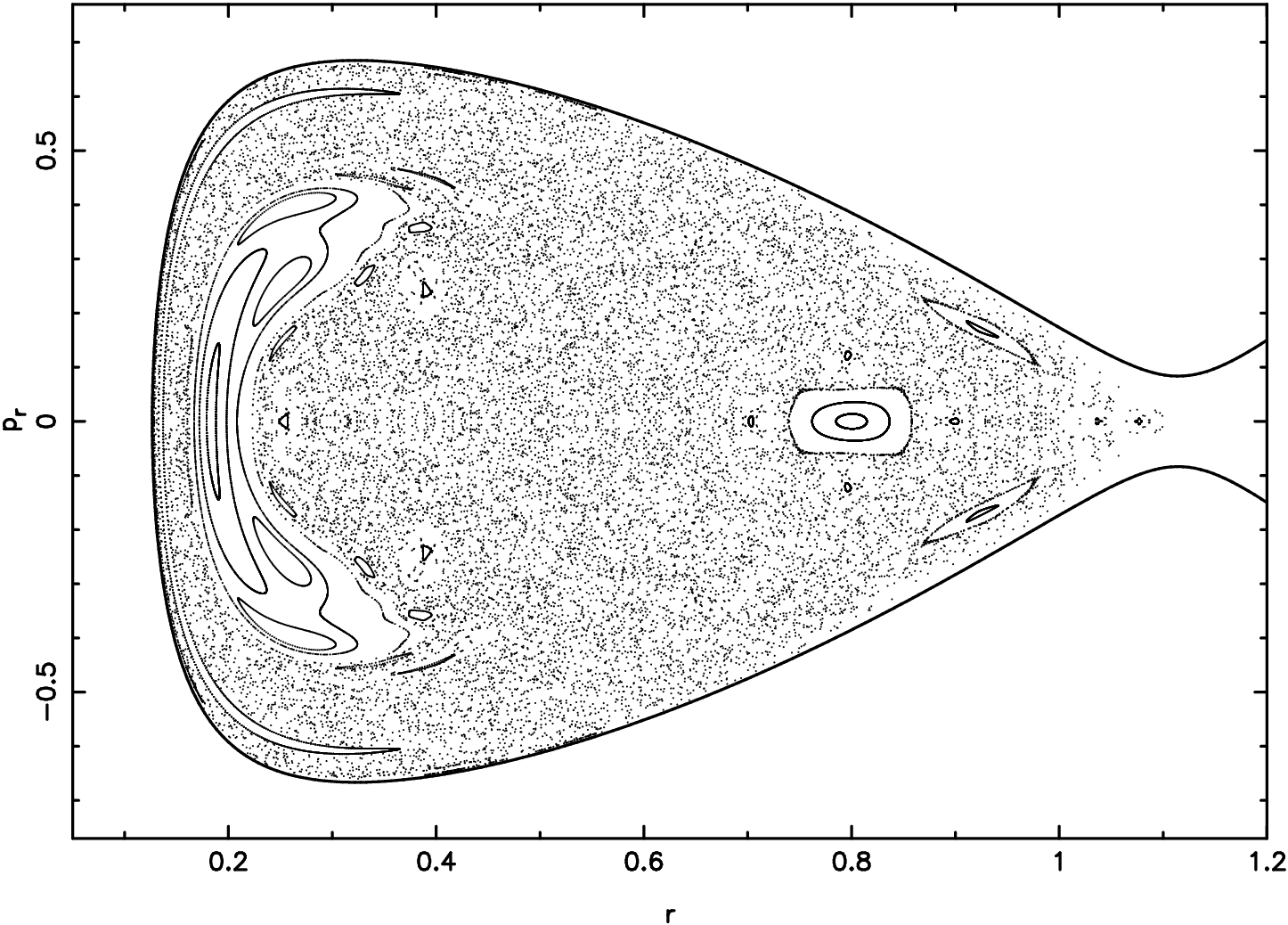}}\hspace{5cm}
                      \rotatebox{270}{\includegraphics*{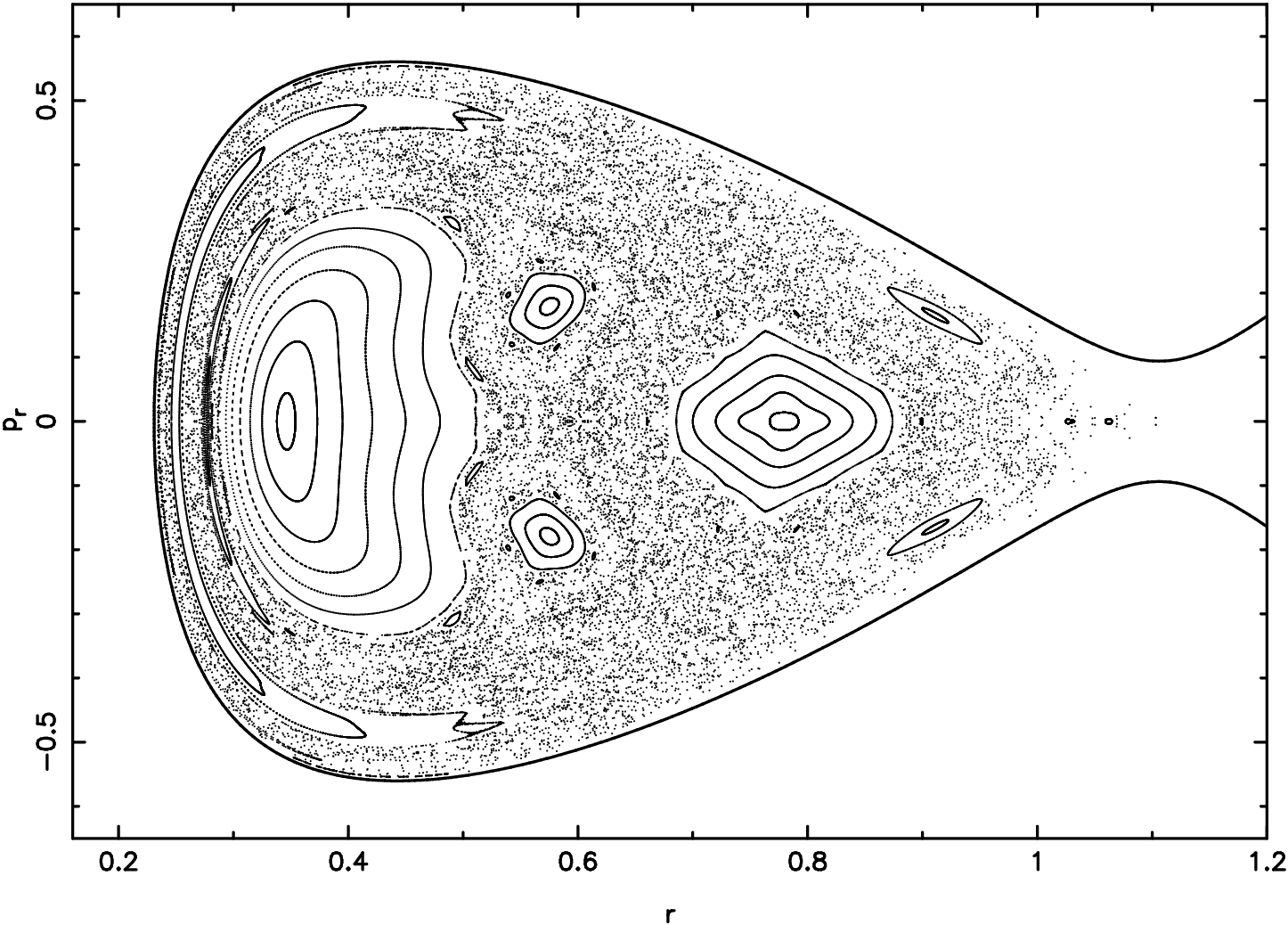}}}
\resizebox{\hsize}{!}{\rotatebox{270}{\includegraphics*{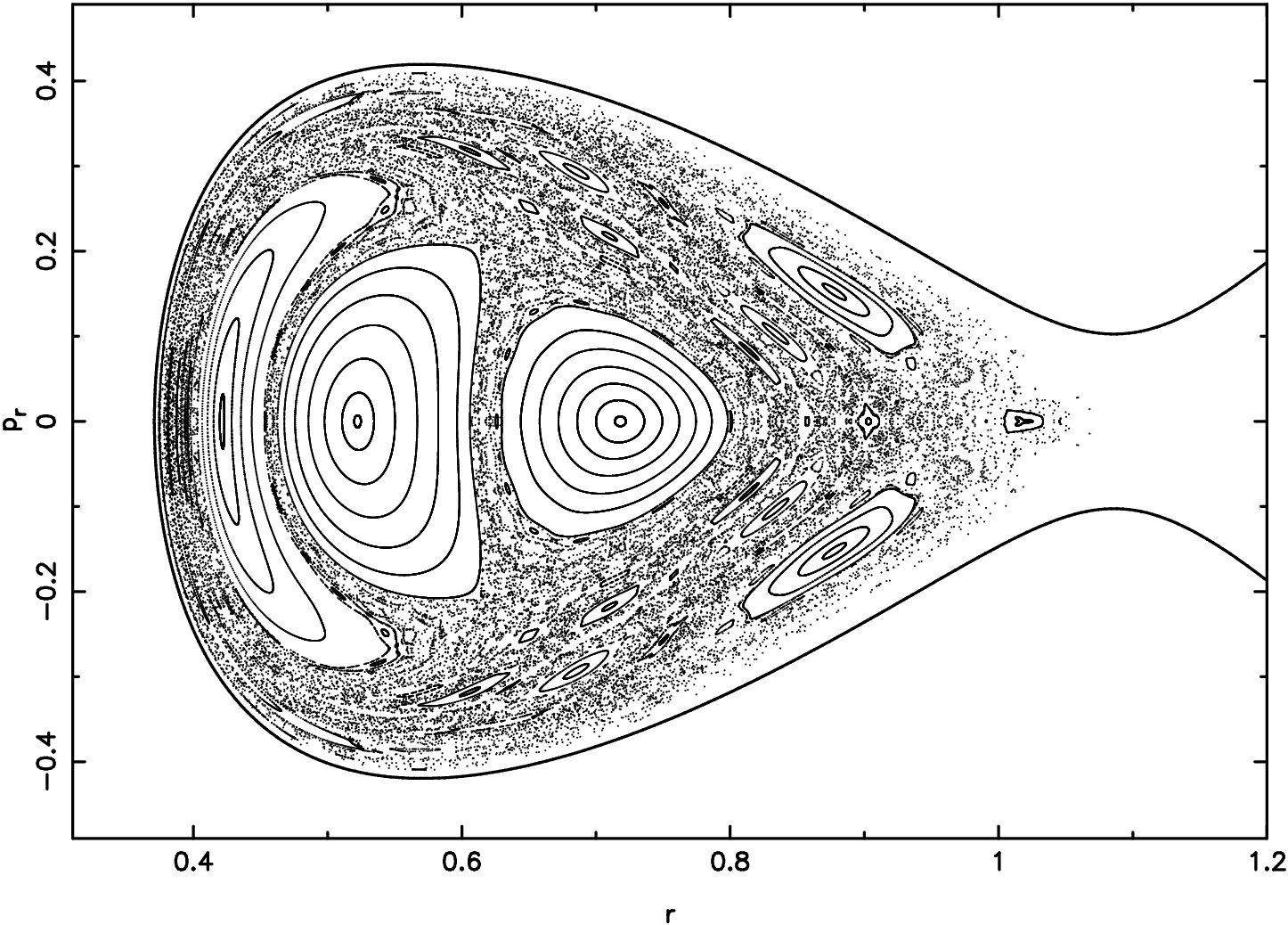}}\hspace{5cm}
                      \rotatebox{270}{\includegraphics*{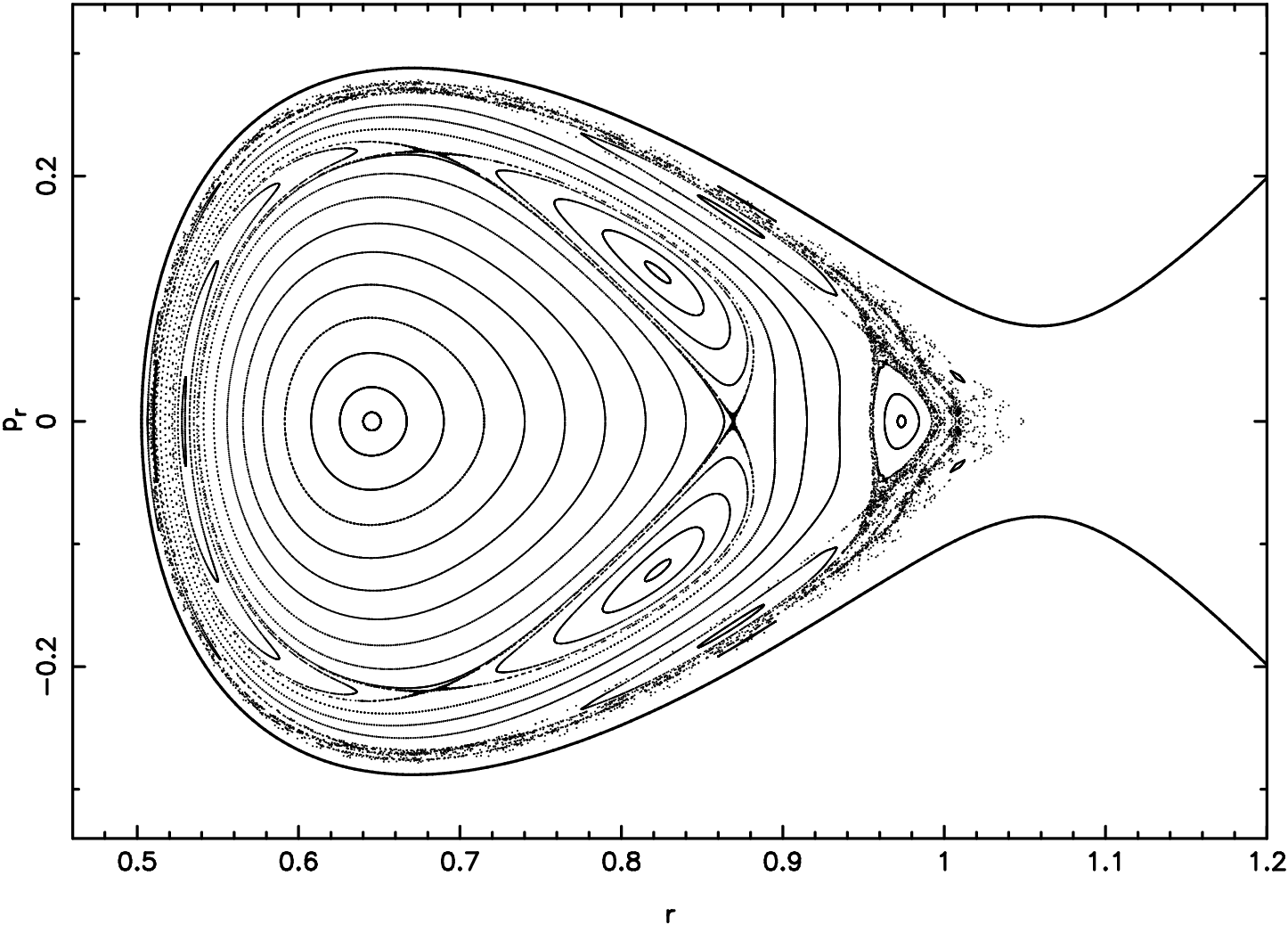}}}
\vskip 0.1cm
\caption{(a-d): The $\left(r,p_r\right)$, $z=0, p_z>0$ phase plane, for the Hamiltonian (4) when (a-upper left): $h=0.32$ and $L_z=0.10$, (b-upper right): $h=0.33$ and $L_z=0.18$, (c-lower left): $h=0.35$ and $L_z=0.28$ and (d-lower right): $h=0.37$ and $L_z=0.36$.}
\end{figure*}

Figure 4a-d shows the $\left(r,p_r\right)$, $z=0, p_z>0$ surfaces of section for the Hamiltonian (4), obtained by numerical integration of the equations of motion (3). In all cases the outermost solid curves, which are the ZVCs, are open. The values of the energy $h$ and the angular momentum $L_z$, are chosen as we have described in the previous section. In Figure 4a, we have the case where $h=0.32$ and $L_z=0.10$. As one can see the majority of the phase plane is covered by chaotic orbits, producing a unified chaotic sea, while there are also areas of regular motion. In addition to the above, one can also observe smaller islands of invariant curves, embedded in the chaotic sea, which are produced by secondary resonant orbits. Figure 4b shows the $\left(r,p_r\right)$ phase plane when $h=0.33$ and $L_z=0.18$. In this case the area on the phase plane occupied by regular orbits is larger than in Fig. 4a, while the resonant islands look more prominent. Figure 4c shows the $\left(r,p_r\right)$ phase plane when $h=0.35$ and $L_z=0.28$. Here the extent of the chaotic area has decreased and the structure of the phase plane is very complicated and interesting as well. We observe that apart from the invariant curves corresponding to the basic families of orbits of the dynamical system, they have appeared numerous small embedded islands of invariant curves, produced by complicated secondary resonant orbits. We will study these secondary resonant orbits later in the next Section. Figure 4d shows the $\left(r,p_r\right)$ phase plane when $h=0.37$ and $L_z=0.36$. In this case, things are quite different than in the previous phase planes. Here the majority of orbits are regular, while a very small chaotic layer which is confined in the outer parts of the phase plane is present. Secondary resonances are still observed.
\begin{figure}[!tH]
\centering
\resizebox{\hsize}{!}{\rotatebox{0}{\includegraphics*{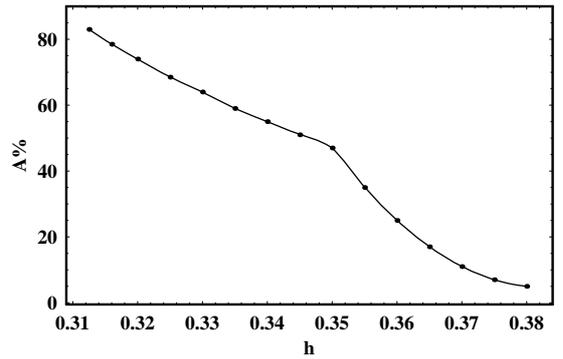}}}
\caption{A plot of the relationship between the percentage of the area $A\%$ covered by chaotic orbits in the $\left(r,p_r\right)$ phase plane and the value of the energy $h$ of the dynamical system.}
\end{figure}

The main conclusions from the above analysis on the $\left(r,p_r\right)$ phase planes of the dynamical system are: (i) The area on the phase planes occupied by regular orbits, increases significantly as the value of the energy increases. At the same time, the chaotic regions are reduced with the increase of the value of the energy $h$. Our numerical calculations indicate that the area corresponds to regular orbits in the $\left(r,p_r\right)$ phase planes, increases as we proceed to larger values of $h$ and $L_z$, always following relation (7) and the same way of choosing the pairs $\left[h, L_z\right]$ as has been described in the previous Section. (ii) As the value of the energy $h$ is increases, we observe that the maximum velocity $p_r$ in the $\left(r,p_r\right)$ phase planes decreases. Furthermore, the whole area of the phase planes is reduced and becomes smaller as the value of the energy $h$ increases. (iii) The relative width of the channel in the ZVC through which orbits can escape to infinity, increases as the values of the energy $h$ and angular momentum $L_z$ increase. Therefore, one may conclude that as we go to larger values of $h$ and $L_z$, the probability of a star to escape from the dynamical system increases.

Figure 5 shows a plot of the relationship between the percentage of the area $A\%$ covered by chaotic orbits in the $\left(r,p_r\right)$ phase plane and the energy of the dynamical system $h$. We see that the percentage $A\%$ decreases as the value of the energy $h$ increases. Note that for small values of $h$, the slope of the curve in the diagram is small, while for larger values of the energy $\left(h \geq 0.35 \right)$ the slope is high. We must point out, that the percentage $A\%$ is calculated as follows: we choose 1000 orbits with different and random initial conditions $\left(r_0, p_{r0}\right)$ in each $\left(r,p_r\right)$ phase plane and then we divide the number of those who produce chaotic orbits to the total number of tested orbits (see Zotos 2011a, 2011b). Here we must remember that the higher is the $h$, the higher is the $L_z$ (see Fig. 1). Therefore, we have a different slope for small values of the the angular momentum $L_z$. This phenomenon is similar to that observed in studies connecting chaos to angular momentum (e.g., Caranicolas \& Innanen 1991; Caranicolas \& Papadopoulos 2003a; Caranicolas \& Zotos 2010; Caranicolas \& Zotos 2011b; Zotos 2011a, 2011b).

\section{Analysis of trapped and escaping orbits}

Let us now proceed and study the various families of the regular orbits that appear in our dynamical model. Figure 6a-d shows four periodic orbits which represent the four basic families of orbits of this galactic system. By the term basic orbits of the dynamical system, we mean resonant orbits of low multiplicity (such as the orbits shown in Fig. 6a-d). Figure 6a shows an orbit with initial conditions: $r_0=0.8, z_0=0, p_{r0}=0$ while, the value of $p_{z0}$ is found from the energy integral (4), for all orbits. This orbits is a typical example of the 1:1 resonant. The values of the energy $h$ and the angular momentum $L_z$, are as in Fig. 4a. In Figure 6b wee see a regular orbit, characteristic of the 2:1 resonant. This orbit has initial conditions: $r_0=0.1832, z_0=0, p_{r0}=0$, while the values of the other parameters are as in Fig. 4a. Figure 6c depicts a periodic orbit with initial conditions: $r_0=0.925, z_0=0, p_{r0}=0.169$. This orbit characterize the family of the 2:3 resonant orbits. The values of the energy and the angular momentum for this orbit, are as in Fig. 4a. In Figure 6d we can observe a typical example of a 4:3 resonant orbit, which has initial conditions: $r_0=0.2757, z_0=0, p_{r0}=0$. For this orbit the values of $h$ and $L_z$ are as in Fig. 4b. All orbits were calculated for a time period of 100 time units.
\begin{figure*}[!tH]
\centering
\resizebox{\hsize}{!}{\rotatebox{0}{\includegraphics*{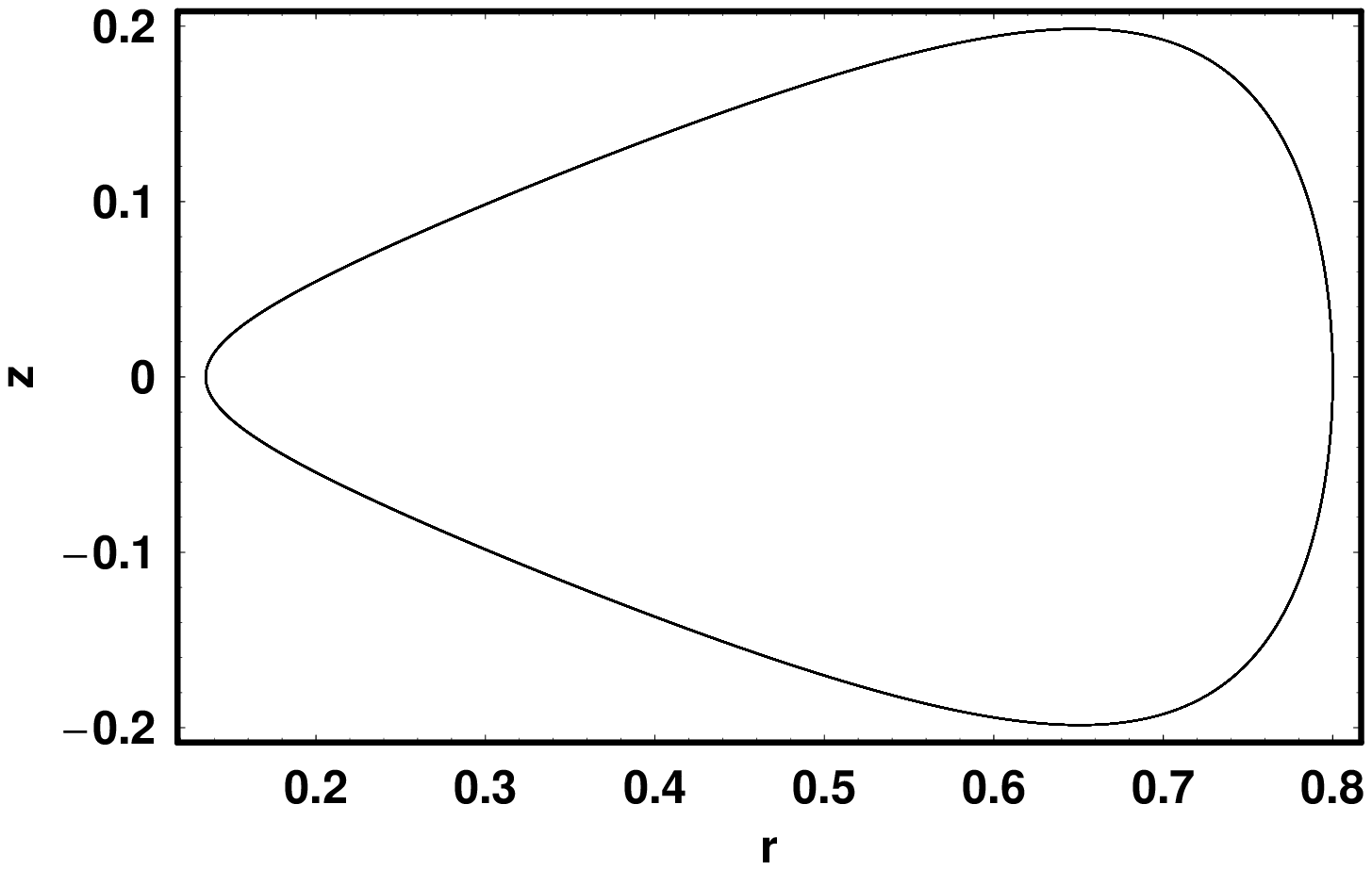}}\hspace{5cm}
                      \rotatebox{0}{\includegraphics*{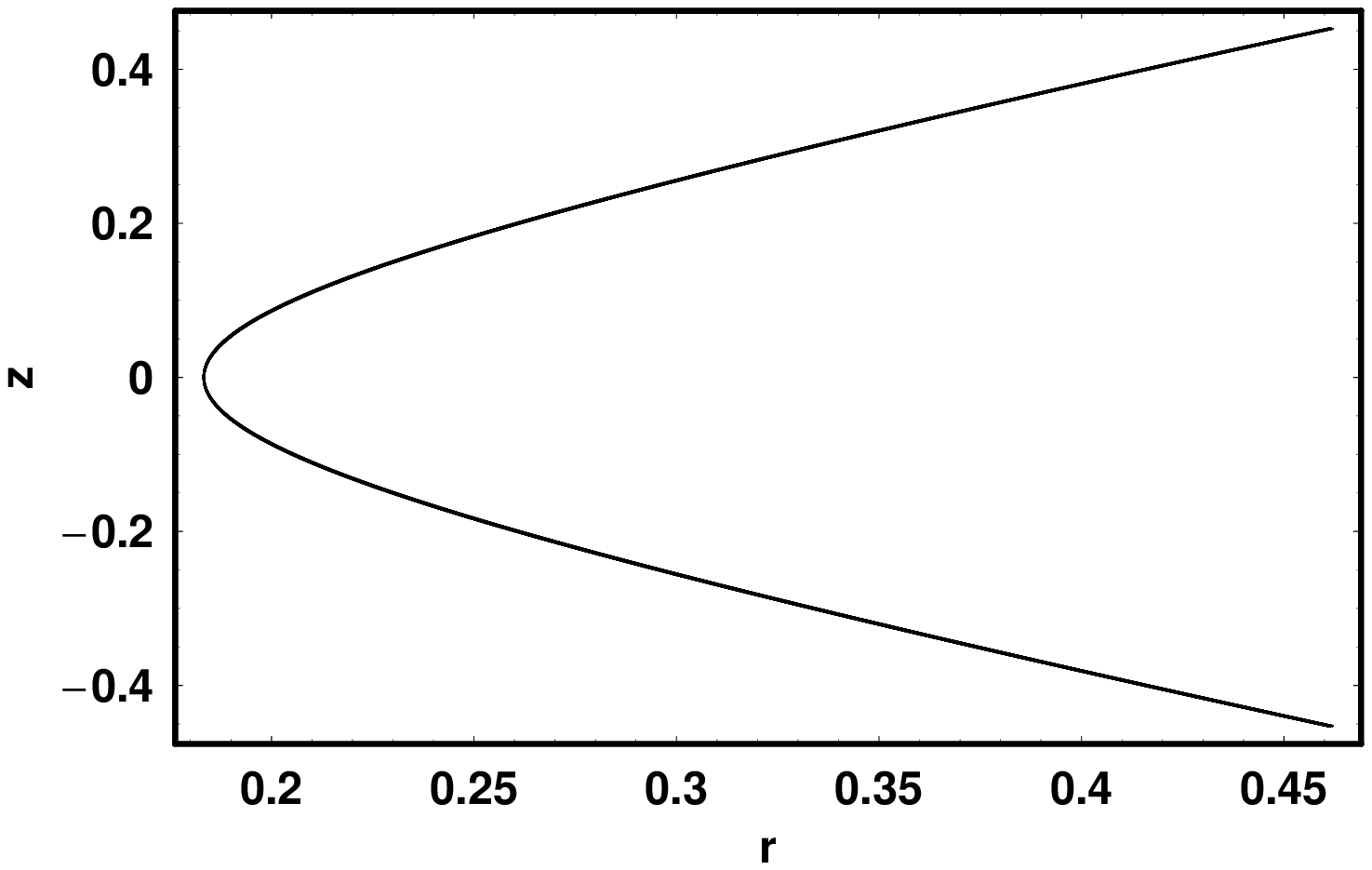}}}
\resizebox{\hsize}{!}{\rotatebox{0}{\includegraphics*{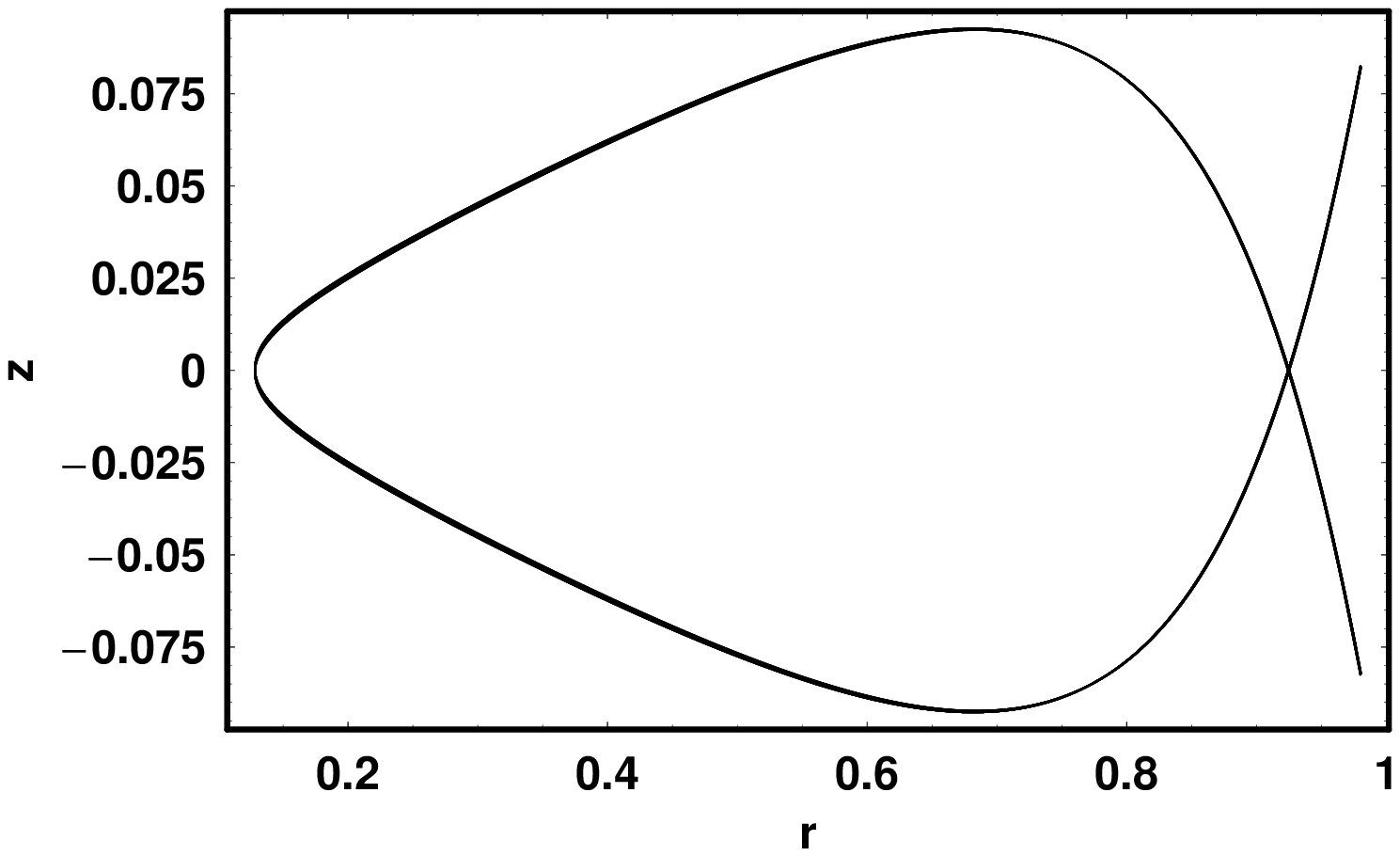}}\hspace{5cm}
                      \rotatebox{0}{\includegraphics*{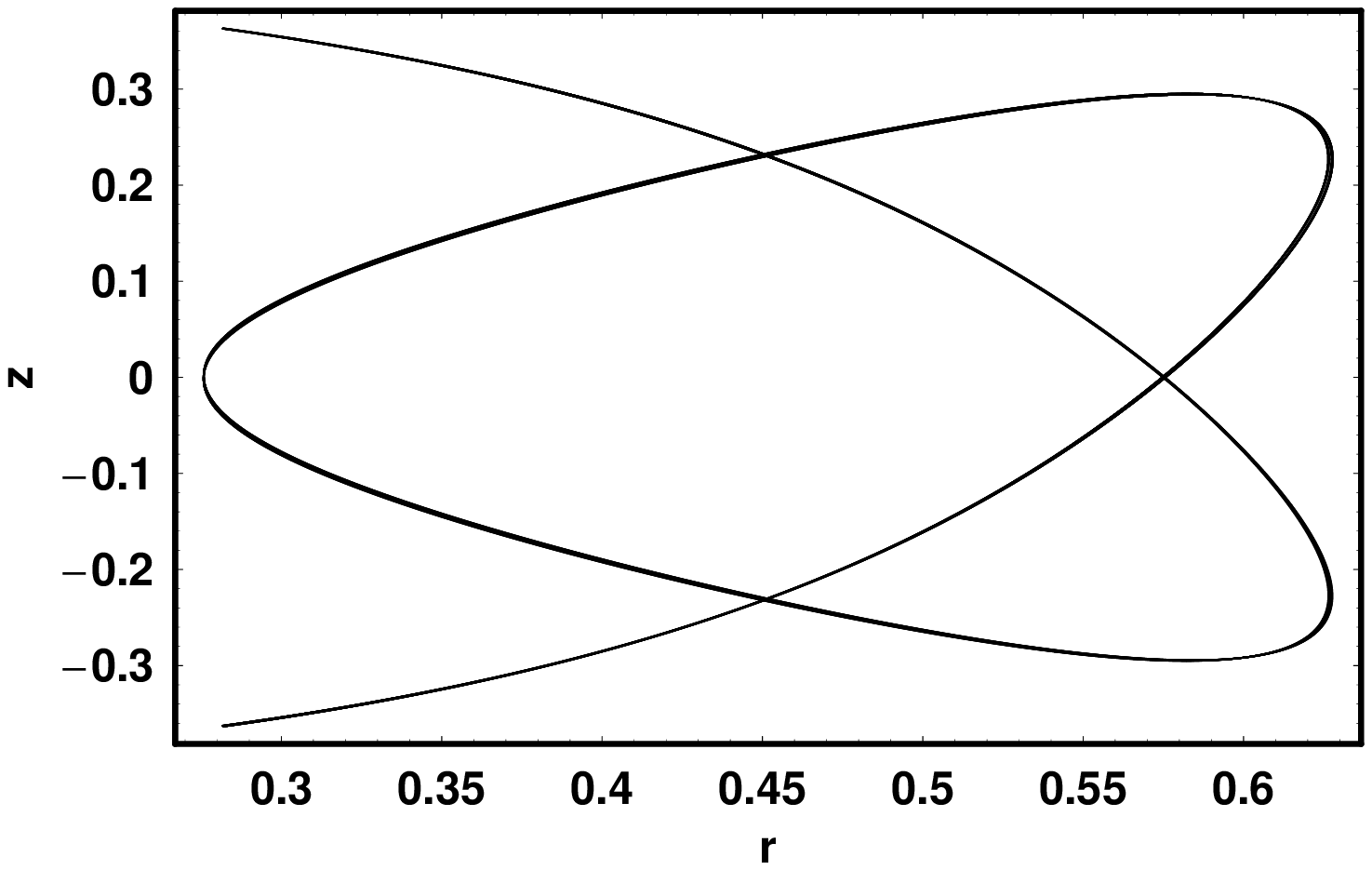}}}
\vskip 0.1cm
\caption{(a-d): Four periodic orbits of the dynamical system. (a-upper left): A 1:1 resonant periodic orbit, (b-upper right): a 2:1 resonant periodic orbit, (c-lower left): a 2:3 resonant periodic orbit and (d-lower right): a 4:3 resonant periodic orbit. More details and the initial conditions of the orbits are given in the text.}
\end{figure*}

Apart from the families of the basic periodic orbits, as we can see from the phase planes shown in Fig. 4a-d, the dynamical system contains numerous and more complicated periodic orbits, produced by secondary resonances. These secondary resonances correspond to multiple islands of invariant curves in the $\left(r,p_r\right)$ phase planes. Figure 7a-h shows eight secondary resonant periodic orbits of the dynamical system. Figure 7a shows a resonant orbit with initial conditions: $r_0=0.3908, z_0=0, p_{r0}=0$, while the values of the energy and the angular momentum are as in Fig. 4c. In Figure 7b we can see a periodic orbit which has initial conditions: $r_0=0.39549, z_0=0, p_{r0}=0$. The values for all the other parameters are as in Fig. 4c. Figure 7c depicts a resonant orbit with initial conditions: $r_0=1.0627, z_0=0, p_{r0}=0$, while the values of $h$ and $L_z$ are as in Fig. 4a. In Figure 7d, one can observe an orbit with initial conditions: $r_0=0.7027, z_0=0, p_{r0}=0$. For this orbit the values of the energy and the angular momentum are as in Fig. 4a. Figure 7e shows a secondary resonant orbit, with initial conditions: $r_0=0.2543, z_0=0, p_{r0}=0$. The values for all the other parameters are as in Fig. 4a. In Figure 7f, one can see a periodic complicated orbit, which has initial conditions: $r_0=0.8574, z_0=0, p_{r0}=0$. The values of $h$ and $L_z$ are as in Fig. 4c. A much more complicated orbit with initial conditions: $r_0=0.1597, z_0=0, p_{r0}=0$, is shown in Figure 7g. The values of the energy and the angular momentum for this orbit are as in Fig. 4a. Finally, in Figure 7h we can observe a resonant periodic orbit which has initial conditions: $r_0=0.39692, z_0=0, p_{r0}=0$, while the values for all the other parameters for this orbit, are as in Fig. 4c. All resonant periodic orbits shown in Fig. 7a-h were calculated for a time period of 100 time units. It is quite interesting to note, that a simple dynamical model as this we use in the present research, is able to produce so many secondary resonant orbits.

Here, we must note that periodic orbits is an important factor of the internal dynamical evolution in galaxies. Romero-G\'{o}mez, et al., (2011) computed the orbits confined by the invariant manifolds of the unstable periodic orbits located at the ends of the bar of the Milky Way. Moreover, the chaotic orbits by the invariant manifolds associated to periodic orbits affect the barred and spiral structure of a galaxy. Furthermore, bars in galaxies are mainly supported by particles trapped around stable periodic orbits. These orbits represent oscillatory motion with only one frequency, which is the bar driving frequency and miss free oscillations. Maciejewski \& Athanassoula, (2007) revealed that a similar situation takes place in double bars: particles get trapped around parent orbits, which in this case represent oscillatory motion with two frequencies of driving by the two bars and which also lack free oscillations. Thus, the parent orbits, which constitute the backbone of an oscillating potential of two independently rotating bars, are the double-frequency orbits. These orbits do not close in any reference frame, but they map on to closed curves called loops. Trajectories trapped around the parent double-frequency orbit map on to a set of points confined within a ring surrounding the loop. The families of periodic orbits in a galaxy, besides of the barred structure may also affect the spiral arms or the thick disks.
\begin{figure*}[!tH]
\centering
\resizebox{\hsize}{!}{\rotatebox{0}{\includegraphics*{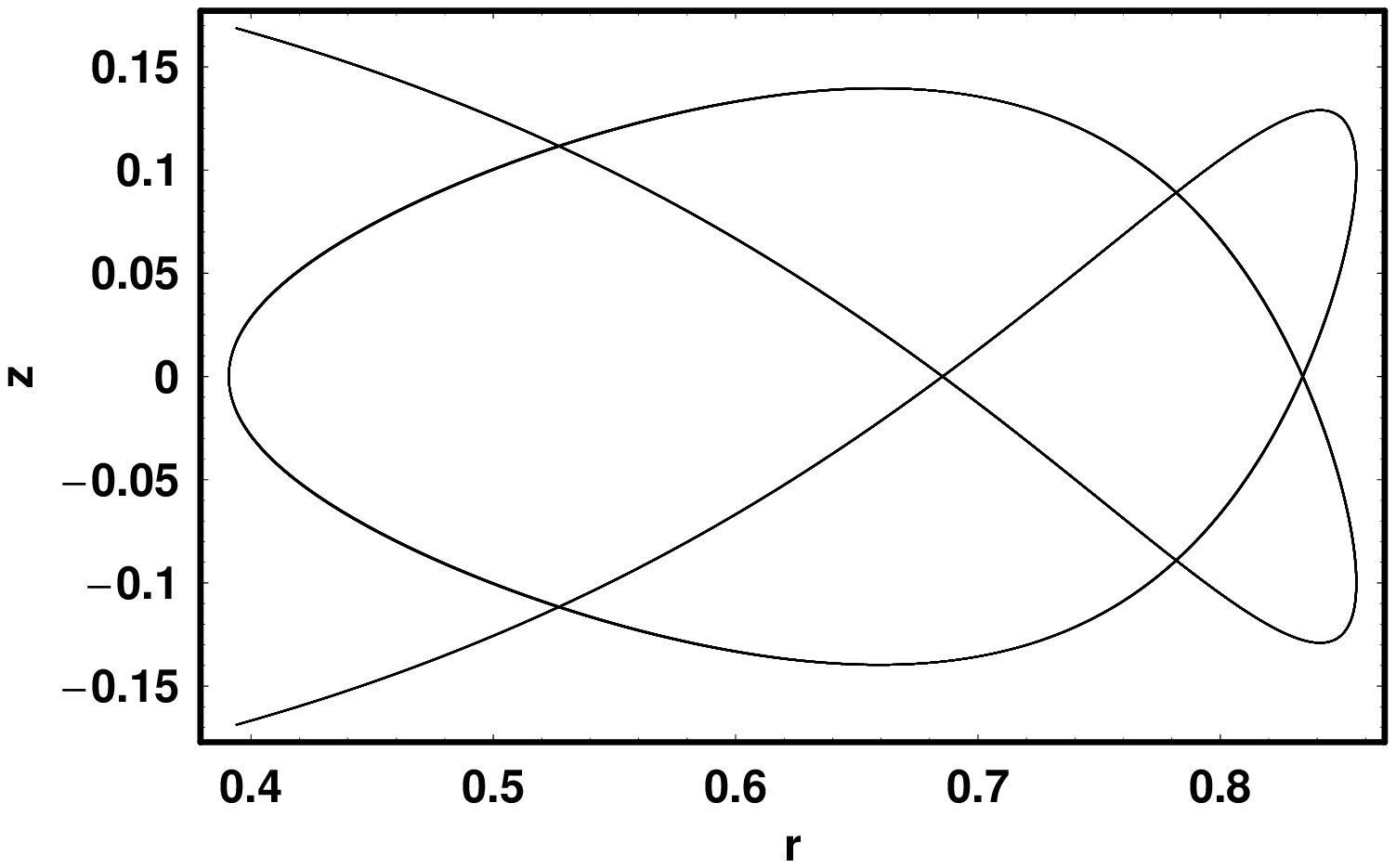}}\hspace{5cm}
                      \rotatebox{0}{\includegraphics*{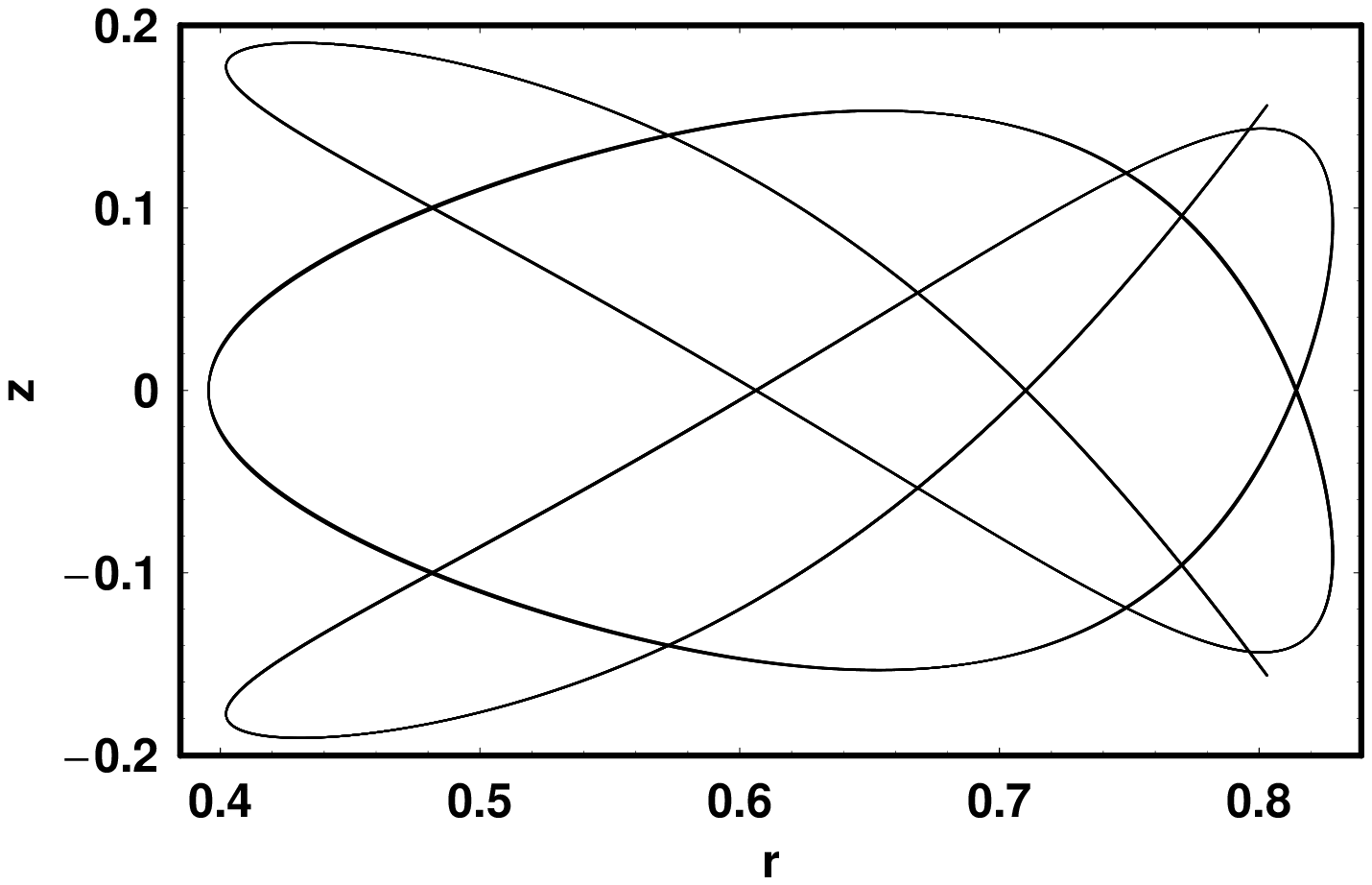}}}
\resizebox{\hsize}{!}{\rotatebox{0}{\includegraphics*{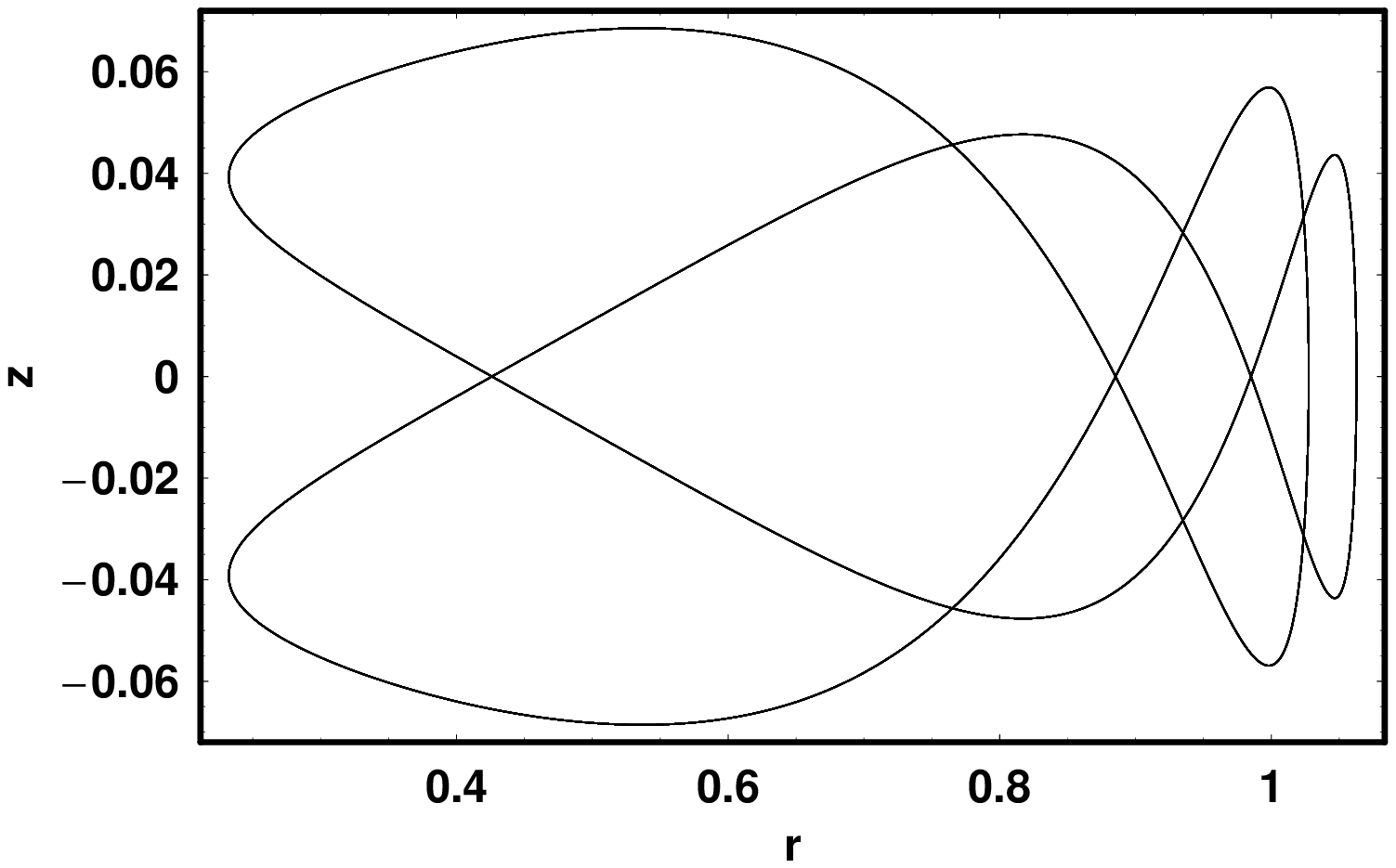}}\hspace{5cm}
                      \rotatebox{0}{\includegraphics*{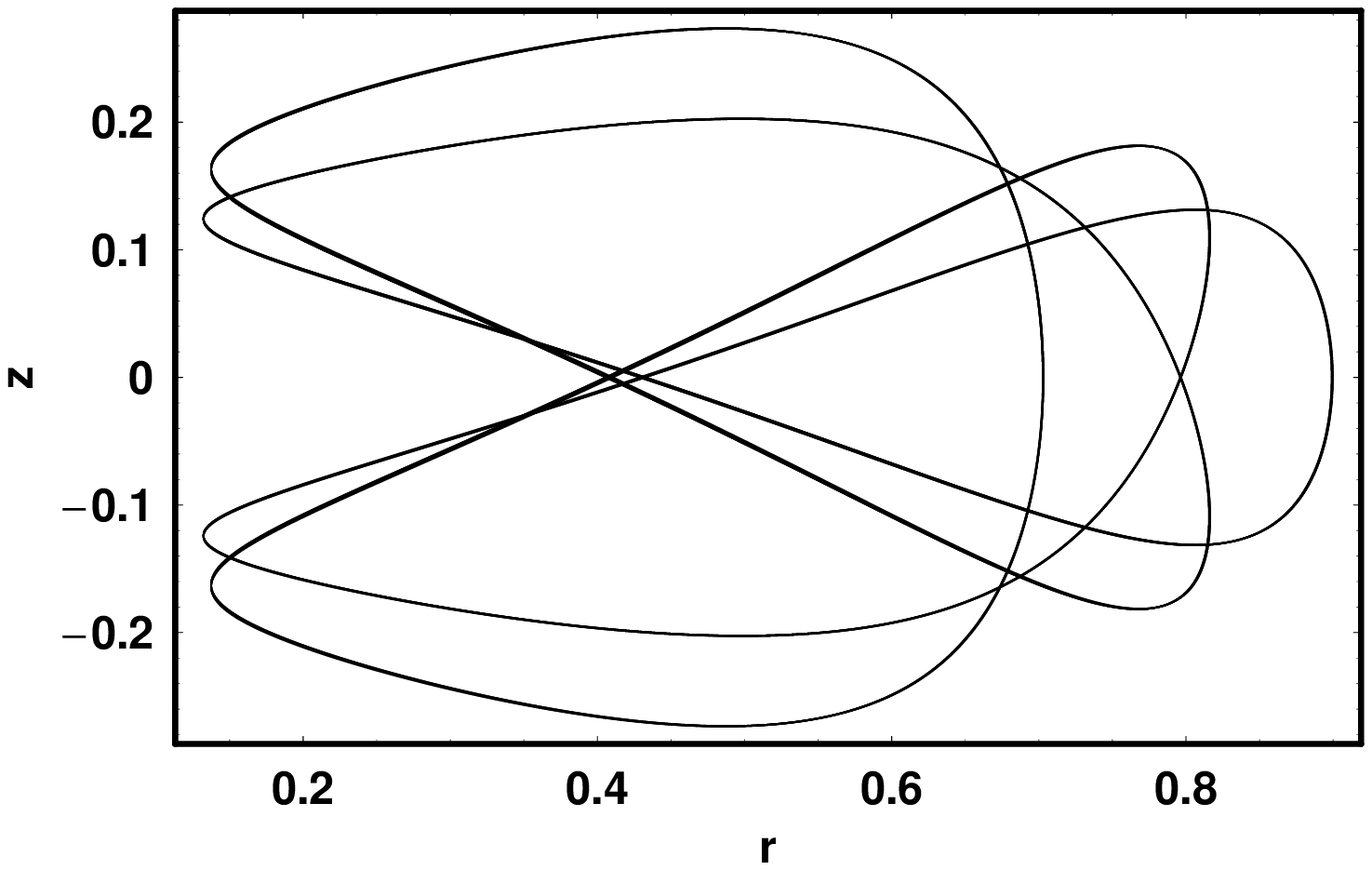}}}
\resizebox{\hsize}{!}{\rotatebox{0}{\includegraphics*{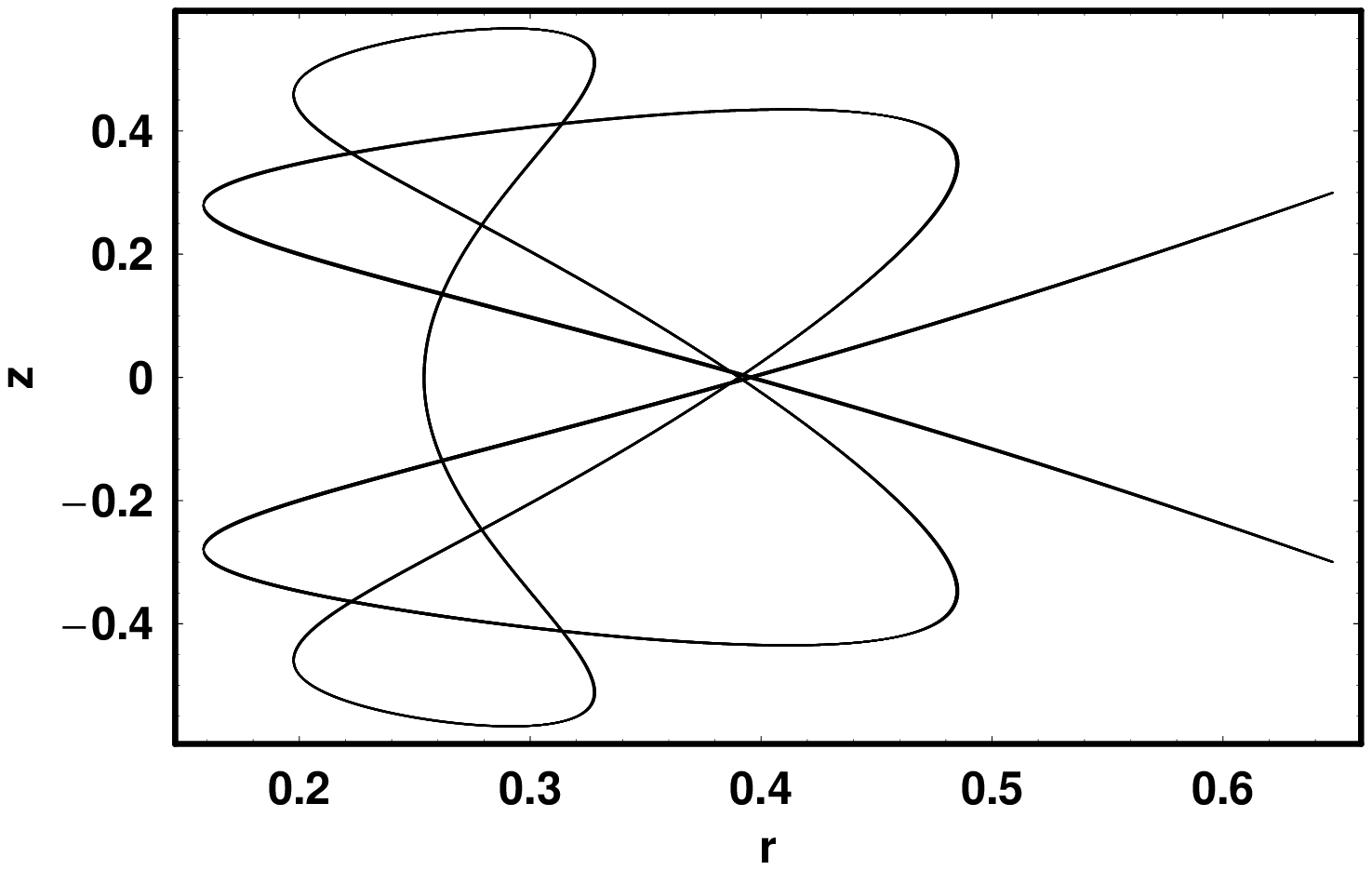}}\hspace{5cm}
                      \rotatebox{0}{\includegraphics*{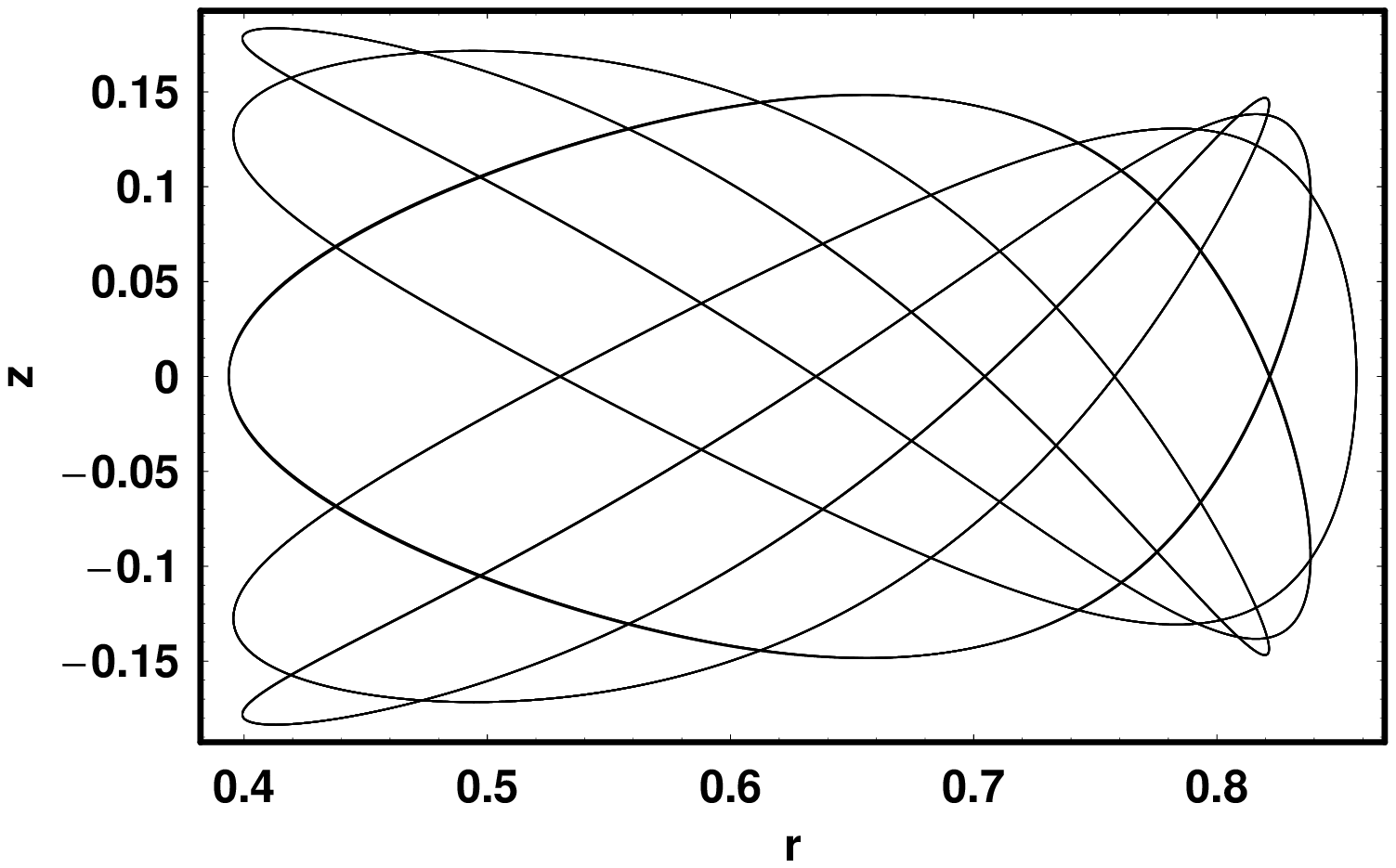}}}
\resizebox{\hsize}{!}{\rotatebox{0}{\includegraphics*{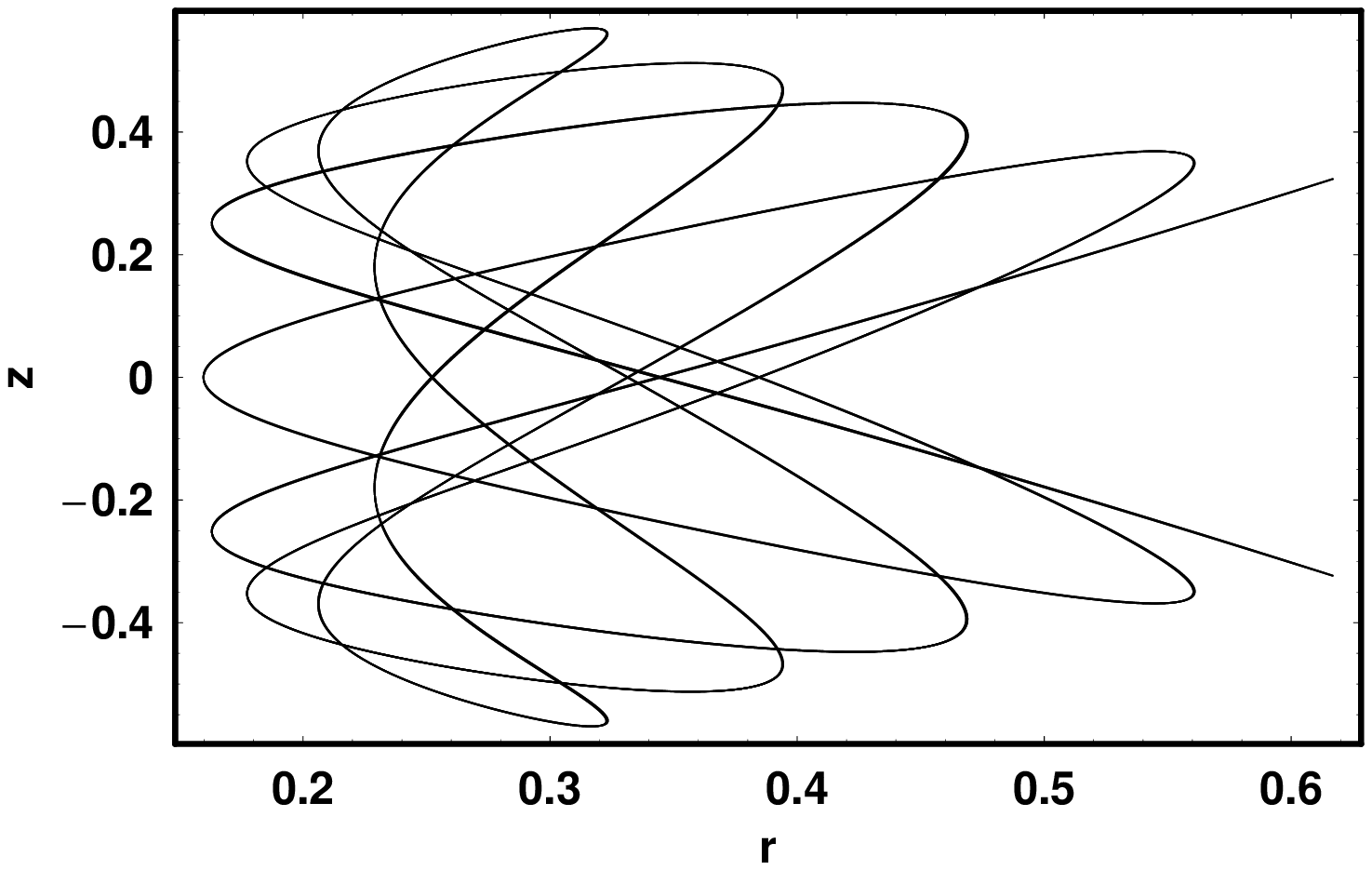}}\hspace{5cm}
                      \rotatebox{0}{\includegraphics*{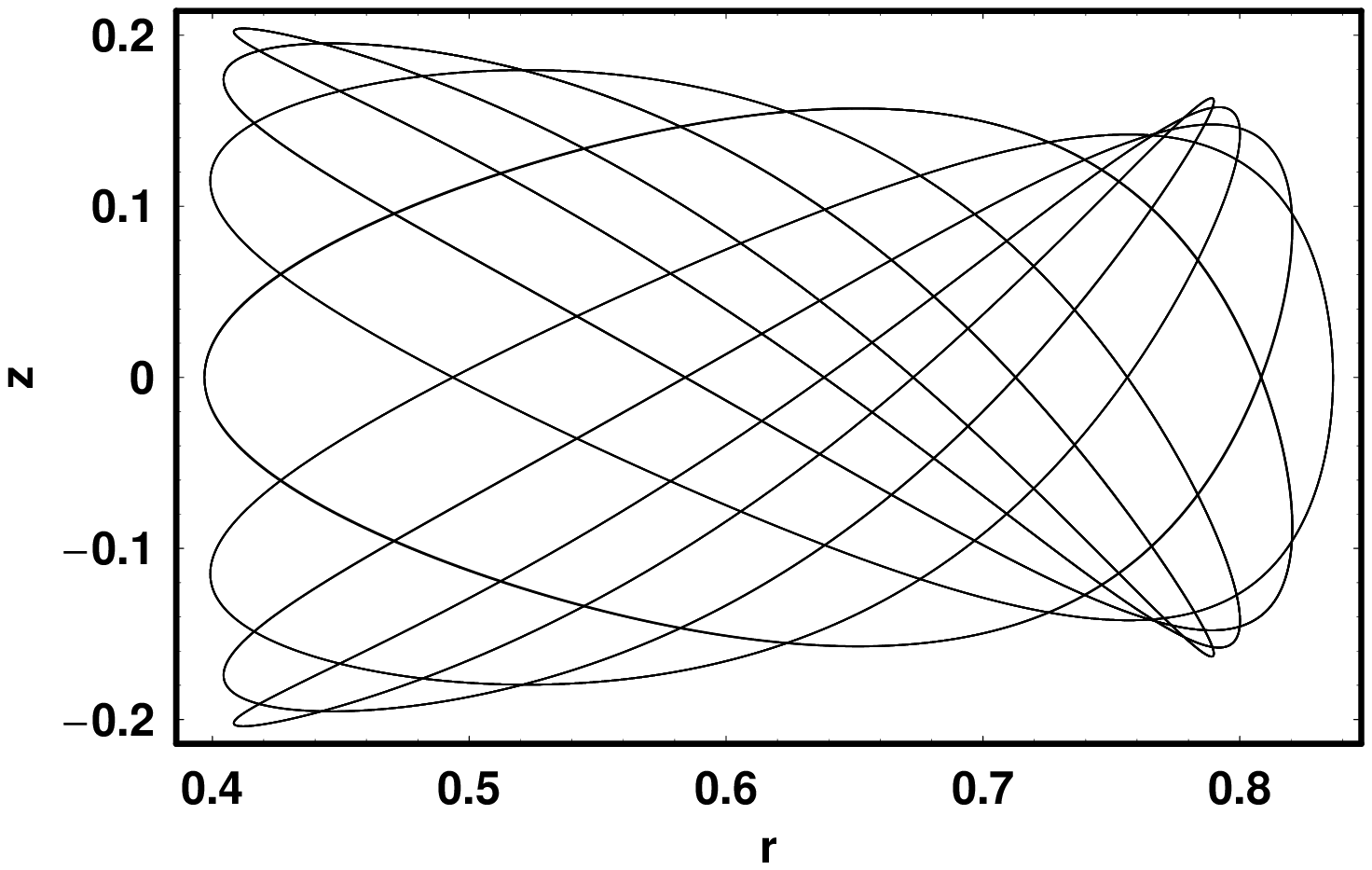}}}
\vskip 0.1cm
\caption{(a-h): Secondary resonant periodic orbits of the dynamical system. The initial conditions of the orbits are given in the text.}
\end{figure*}

Our numerical calculations, suggest that all regular orbits cannot escape from the dynamical system and therefore they are trapped regular orbits. We have tested more than 1000 regular orbits with different and random initial conditions $\left(r_0,z_0=0,p_{r0}\right)$ in each phase plane (for different values of the energy $h$ and the angular momentum $L_z$). In all cases, the regular orbits (basic orbits and also orbits correspond to secondary resonances) are trapped, regardless of the total integration time interval and they remain inside the open ZVC.
\begin{figure*}[!tH]
\centering
\resizebox{\hsize}{!}{\rotatebox{0}{\includegraphics*{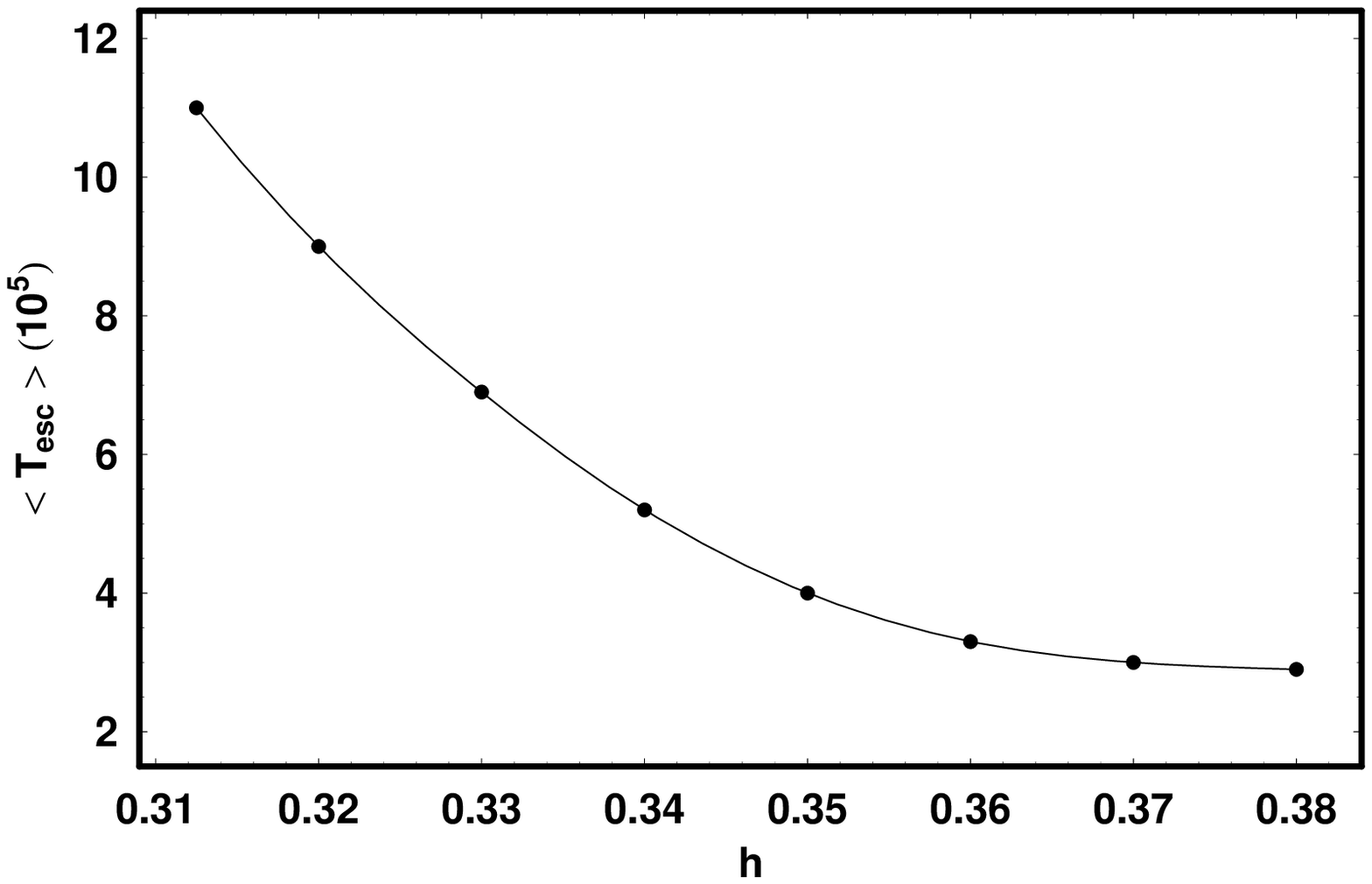}}\hspace{5cm}
                      \rotatebox{0}{\includegraphics*{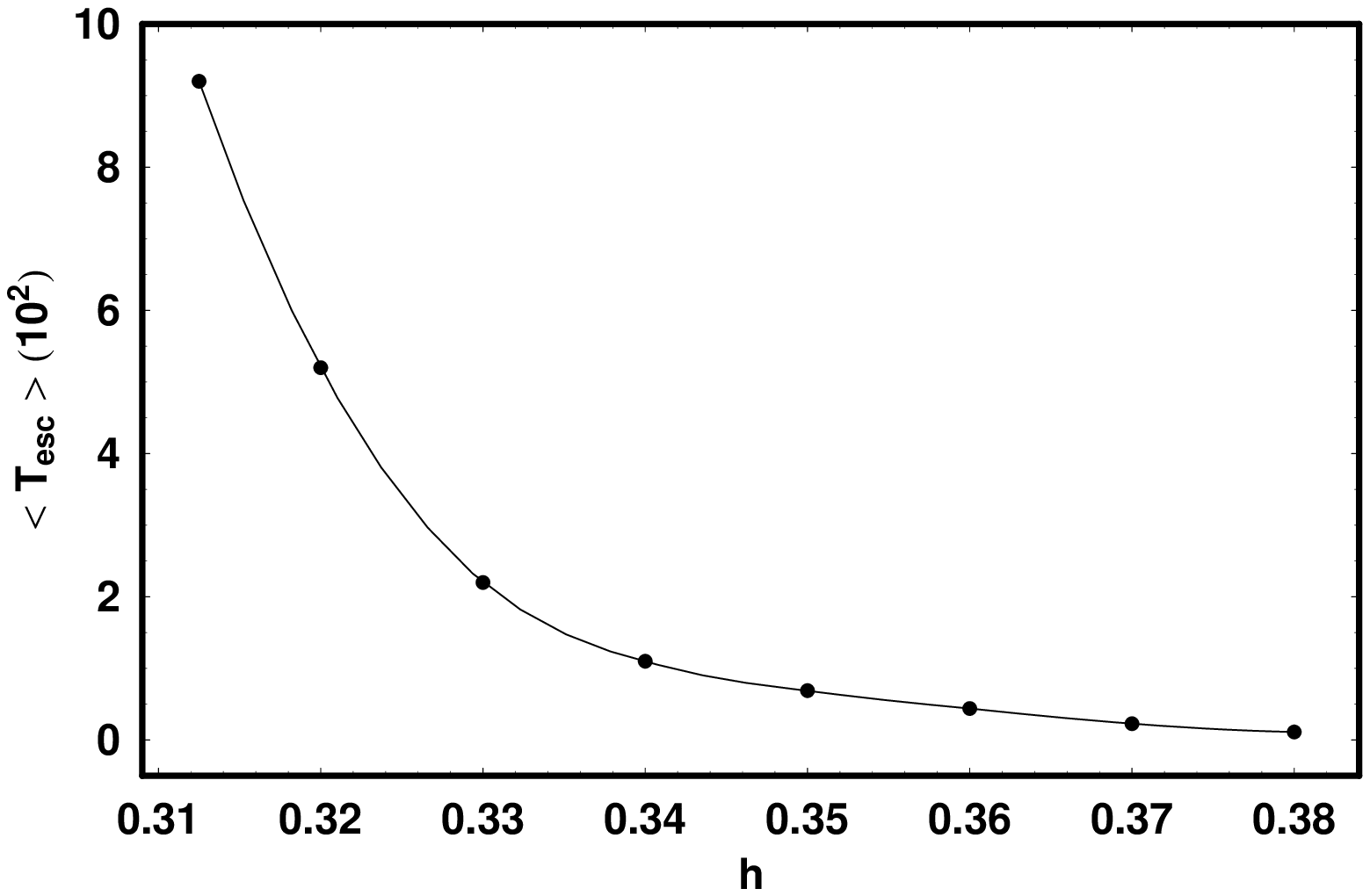}}}
\vskip 0.1cm
\caption{(a-b): The evolution of the mean value of the escape period $< T_{esc} >$ for each value of the energy $h$ for (a-left) the trapped chaotic orbits and (b-right) the fast escaping chaotic orbits.}
\end{figure*}

What is much more interesting, is the behavior of the chaotic orbits of the dynamical system. The unified chaotic sea shown in the $\left(r,p_r\right)$ phase plane of Fig. 4a, is produced by a single chaotic orbit. This orbit has initial conditions: $r_0=0.48, z_0=0, p_{r0}=0$, while the value of $p_{z0}$ is found from the energy integral (4). This orbit escapes, after spending a time interval of more than $9 \times 10^5$ time units, that is $9 \times 10^{12}$yr, until it finds the escape channel to the open ZVC. On the other hand, recent evidence from the Hubble key project and measurements of the fundamental cosmological parameters from WMAP (e.g., Komatsu et al. 2011), give an estimation for the age of the Universe at $\left(1.375 \pm 0.13 \right) \times 10^{10}$yr. As the time needed for the chaotic orbit to escape $T_{esc}$, is much more larger than the age of the Universe, we can consider it as a ``non escaping orbit". Going one step further, we can say that this orbit is an example of a non escaping orbit, that is a trapped chaotic orbit, which is responsible for the phenomenon of ``trapped chaos" in galactic systems with energies larger than the energy of escape. The unified chaotic region shown in the phase plane of Fig. 4b, is also produced by another trapped chaotic orbit. The initial conditions of this orbit are: $r_0=0.57, z_0=0, p_{r0}=0$. This orbit can also be considered as an ``non escaping" orbit, as it escapes to infinity, after spending a total time interval of more than $7 \times 10^5$ time units. Moreover, the chaotic sea shown in the phase plane of Fig. 4c, as also a product of a trapped chaotic orbit, with initial conditions: $r_0=0.625, z_0=0, p_{r0}=0$. In this case, the escape period of this orbit is about $4 \times 10^5$ time units. Similarly, the confined chaotic layer shown in the phase plane of Fig. 4d, is also produced by a trapped chaotic orbit. This orbit has initial conditions: $r_0=0.512, z_0=0, p_{r0}=0$, while its escape period is about $3 \times 10^5$ time units. We observe, that as the value of the energy $h$ increases, the escape period of the trapped orbits decreases, but in all cases the value of $T_{esc}$ is at least about 100 time larger, than the age of the Universe!
\begin{figure*}[!tH]
\centering
\resizebox{\hsize}{!}{\rotatebox{0}{\includegraphics*{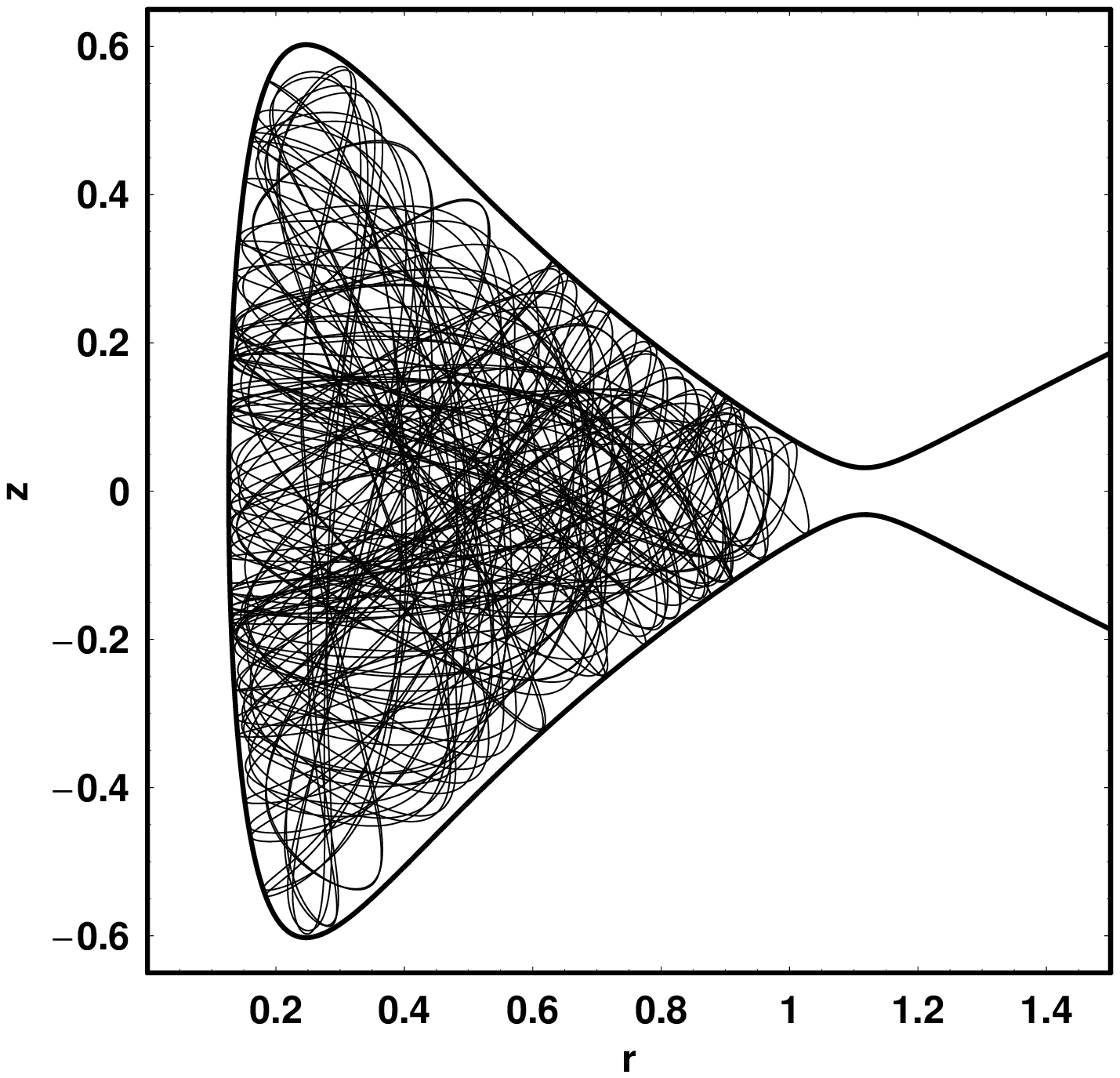}}\hspace{1cm}
                      \rotatebox{0}{\includegraphics*{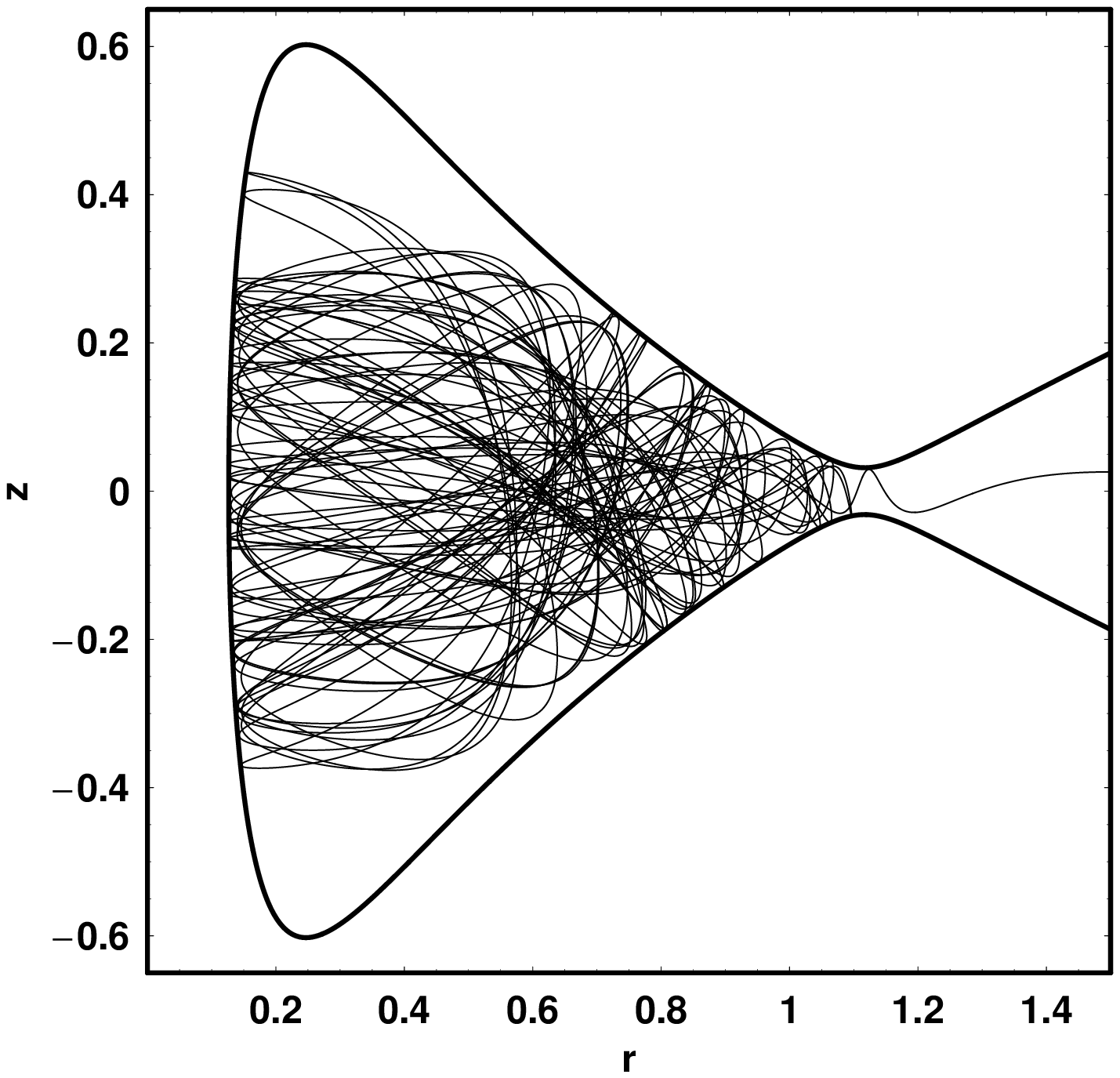}}}
\resizebox{\hsize}{!}{\rotatebox{0}{\includegraphics*{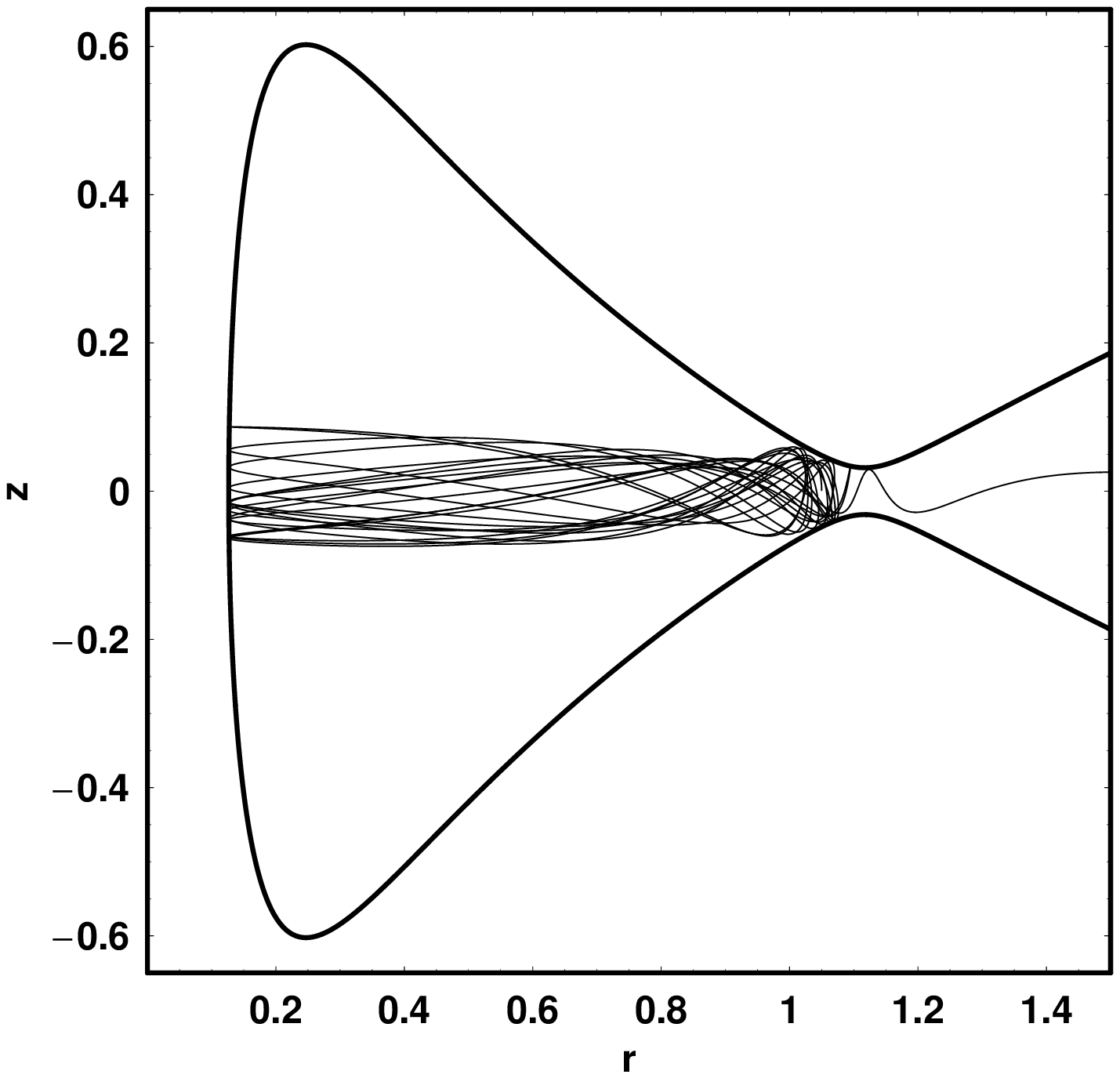}}\hspace{1cm}
                      \rotatebox{0}{\includegraphics*{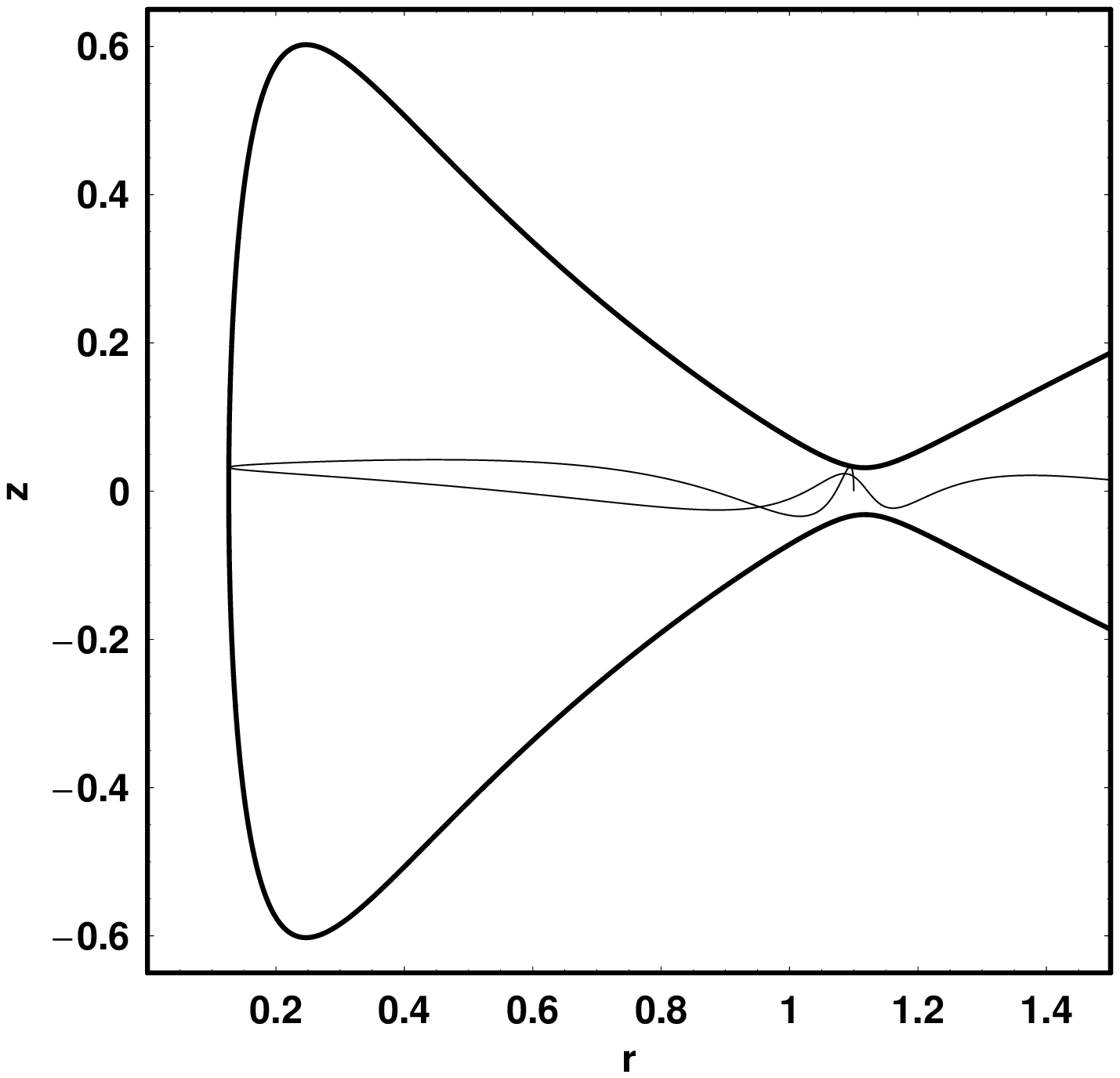}}}
\vskip 0.1cm
\caption{(a-d): Four escaping chaotic orbits of the dynamical system with different periods of escape. (a-upper left): A chaotic orbit which was calculated for 400 time units, (b-upper right): the same chaotic orbit as in Fig. 9a, but for the interval between 1450 and 1700 time units, (c-lower left): another escaping orbit with escape period of 85 time units and (d-lower right): a very fast escaping orbit with an escaping period of 11 time units. The initial conditions of the orbits are given in the text.}
\end{figure*}
\begin{figure*}[!tH]
\centering
\resizebox{\hsize}{!}{\rotatebox{0}{\includegraphics*{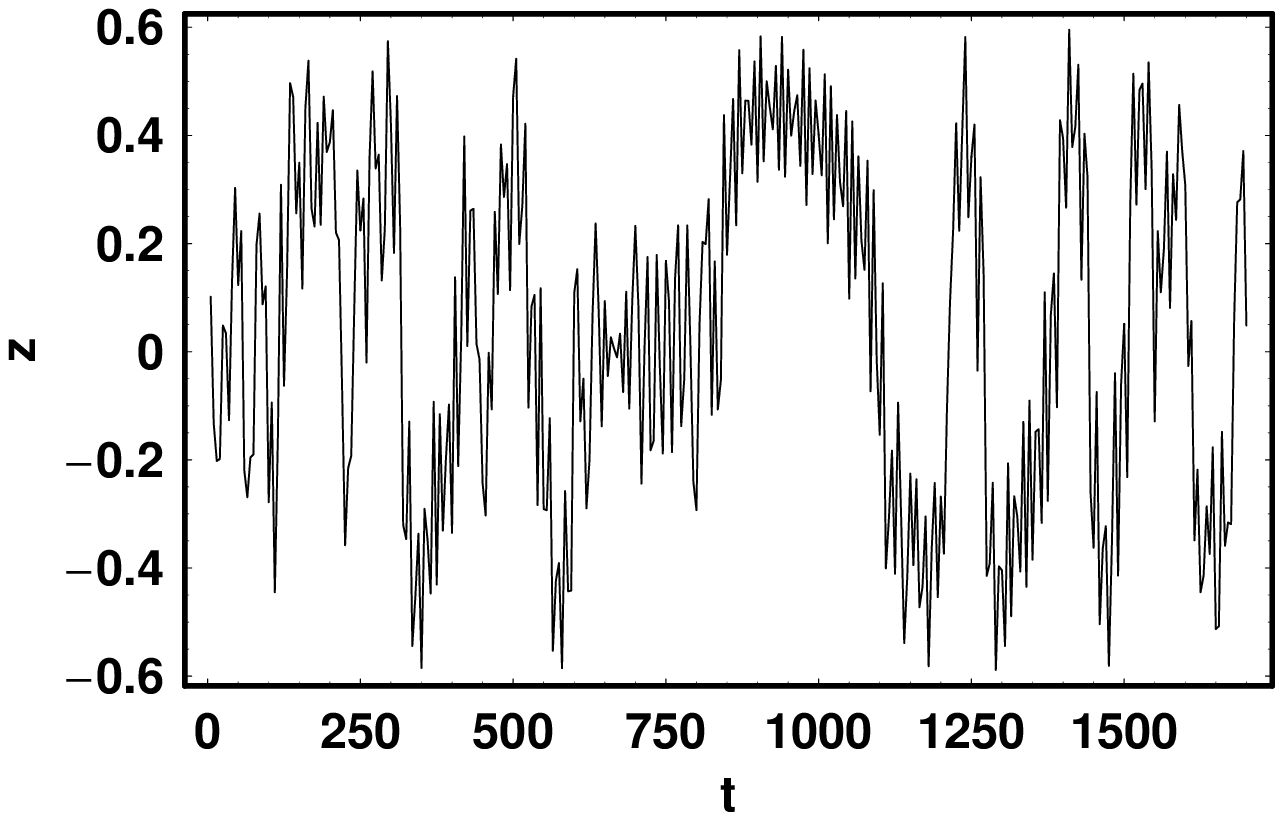}}\hspace{5cm}
                      \rotatebox{0}{\includegraphics*{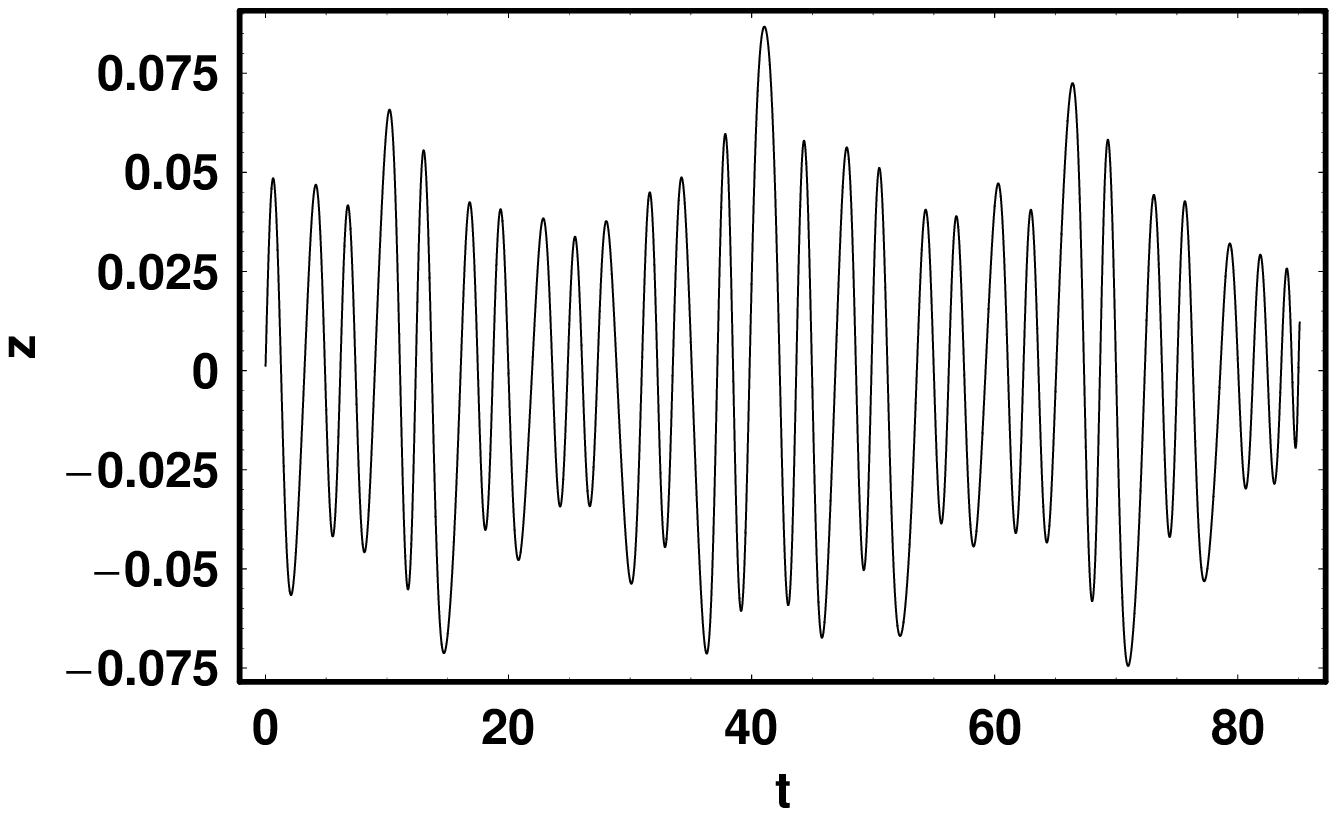}}}
\vskip 0.1cm
\caption{(a-b): Evolution of the star's $z$ height with the time for the orbit shown in (a-left): Fig. 9a and (b-right) Fig. 9c.}
\end{figure*}

Extensive numerical calculations, suggest that about $62\%$ of the total tested chaotic orbits stay inside the ZVC before escaping, for time intervals larger than the age of the Universe. In order to study this behavior in more detail, we have checked the escaping period $T_{esc}$ for 1000 orbits for seven pairs of values: $h=h_{esc}, L_z=L_{zc}- \Delta L$ with $\Delta L=0.01$, that is 7000 orbits. Therefore, we can say that our numerical experiments indicate that about $62\%$ of the chaotic orbits are trapped chaotic orbits. Figure 8a shows a plot describing the evolution of the mean value of the escape period $< T_{esc} >$ for each value of the energy $h$. In this case, the mean values of the escape periods regard the trapped orbits which represent the $62\%$ of the total chaotic orbits. One can observe, that the mean value of the escape period decreases rapidly as the value of the energy increases. Moreover, for large values of the energy $\left(h \geq 0.35 \right)$ the slope is small and the curve tends to become horizontal. As we have discussed earlier, one single orbit covers the entire chaotic domain in each panel of Fig. 4. However, if we start from a different initial condition $\left(r_0,p_{r0}\right)$ in the chaotic region of each phase plane, we cover again the entire phase plane but there is a difference. All the chaotic orbits with different initial conditions, cover each one the entire phase plane but not all of them have the same escape period. This is the basic reason why we have estimated the mean value of the escape period in each case. Moreover, our calculations revealed, that the rest $38\%$ of the total tested chaotic orbits, escape from the open ZVC within very short time intervals. These are fast escaping chaotic orbits. In Figure 8b we can see a plot depicting the evolution of the mean value of the escape period $< T_{esc} >$ for each value of the energy $h$. In this case, the mean values of the escape periods regard the fast escaping orbits which represent the $38\%$ of the total chaotic orbits. One can observe, that the mean value of the escape period decreases much more rapidly than in Fig. 8a, as the value of the energy $h$ increases. Furthermore, for values of the energy $\left(h \geq 0.34 \right)$ the slope is very small and the fitting curve tends to become horizontal. The main conclusion from our orbital analysis, is that as the initial conditions $\left(r_0,z_0=0,p_{r0}\right)$ of an orbit approach increasingly the critical boundary of the Lyapunov orbit, the value of the escape period $T_{esc}$ decreases dramatically.
\begin{figure*}[!tH]
\centering
\resizebox{\hsize}{!}{\rotatebox{0}{\includegraphics*{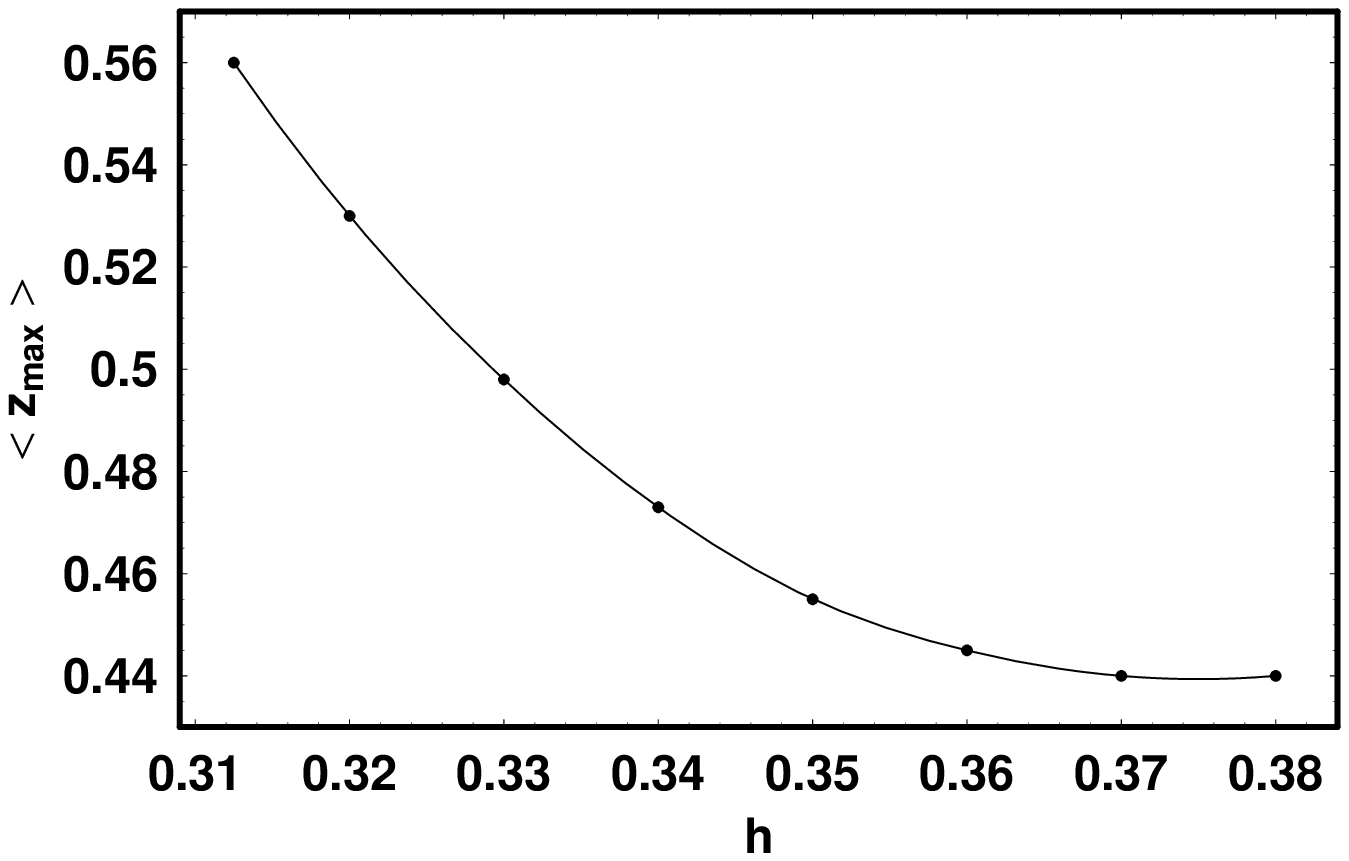}}\hspace{5cm}
                      \rotatebox{0}{\includegraphics*{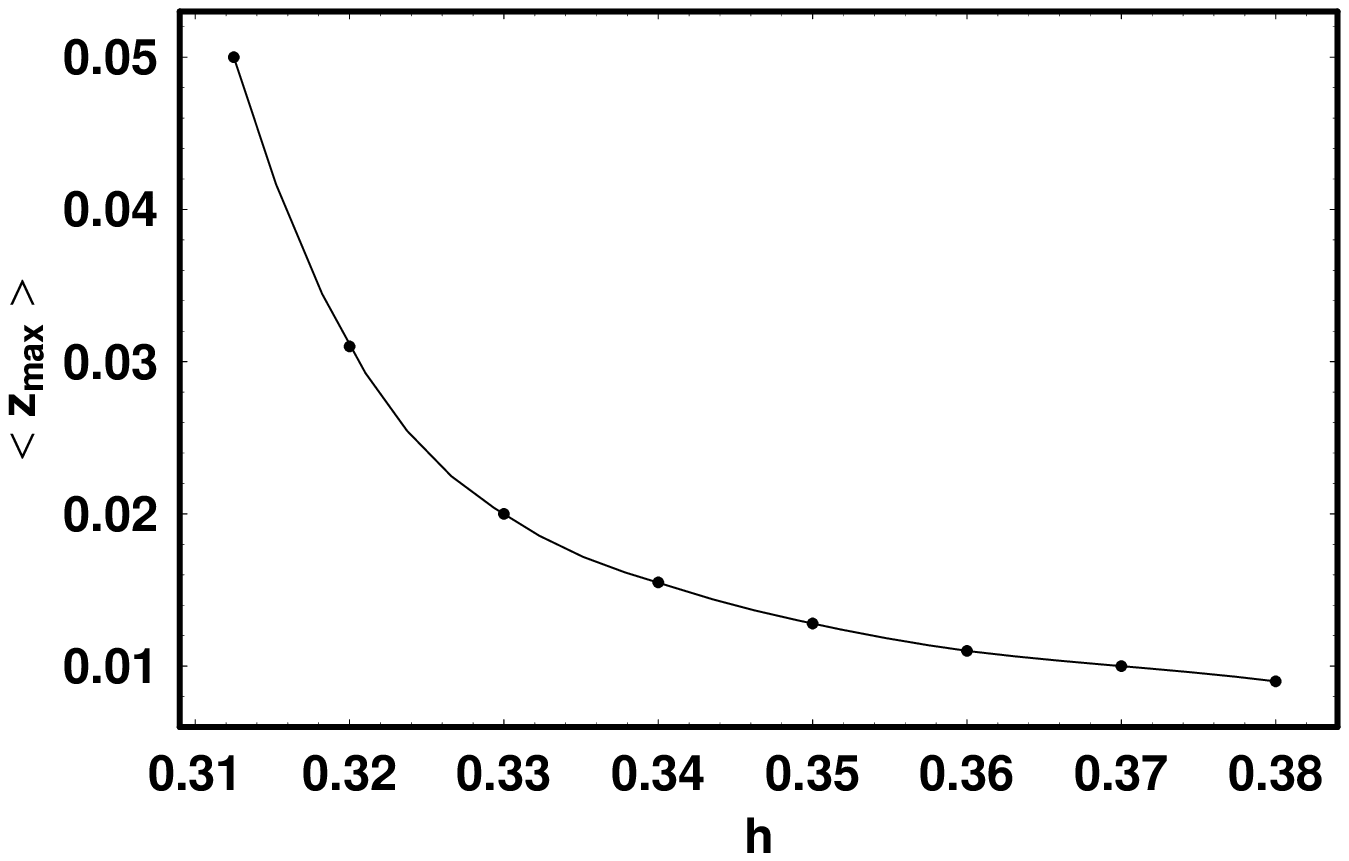}}}
\vskip 0.1cm
\caption{(a-b): The evolution of the average value of the $z$ component $< z_{max} >$ for each value of the energy $h$ for (a-left) the trapped chaotic orbits and (b-right) the fast escaping chaotic orbits.}
\end{figure*}

Figure 9a-d shows four escaping chaotic orbits with different escape periods. Figure 9a shows a chaotic orbit with initial conditions: $r_0=0.92, z_0=0, p_{r0}=0$, while the value of $p_{z0}$ is always found from the energy integral (4). This orbit was calculated for a time period of 400 time units. As we can see, the orbit fills almost all the area inside the ZVC but it does not escape. This orbit escapes after $T_{esc}=1700$ time units. Figure 9b, shows the same orbit as in Fig. 9a bur here the orbit was plotted for the the last 250 time units, that is the interval between 1450 and 1700 time units. In this case, the orbit after spending 1700 time units inside the ZVC, finally finds the channel and escapes to infinity. In Figure 9c we can observe one more escaping orbit, with initial conditions: $r_0=1.05, z_0=0, p_{r0}=0$. This orbit was calculated for only 85 time units. The escape period of this orbit is considerably small. Perhaps we can give an explanation for this behavior, by looking the shape of the orbit given in Fig. 9c. This orbit stays very close to the galactic plane with small values of $z$ (see Fig. 10b). Moreover, the orbit is almost parallel to the channel of escape and therefore, the test particle can find very quickly its way out from the ZVC. Figure 9d shows a very fast escaping chaotic orbit, with initial conditions: $r_0=1.1, z_0=0, p_{r0}=0$. This orbits escapes from the ZVC after only 11 time units. The values of the energy $h$ and the angular momentum $L_z$ for the orbits shown in Fig. 9a-d are 0.32 and 0.10 respectively.

Figure 10a-b shows the evolution of the $z$ component of the orbits shown in Figs. 9a and 9c. Looking at Fig. 10a, we observe that the test particle (star) spends long time intervals, before it escapes from the ZVC, in high values of $z$. On the contrary, things are quite different in Figure 10b, where we can see that for this orbit the star stays near to the galactic plane, with low values of $z$. Therefore, as the star moves almost parallel to the escape channel, it finds relatively quickly the gap in the open ZVC and as a result its escape period is very small. Thus, it seems that there is a relation between the escape period $T_{esc}$ and the $z$ component of the orbits. In an effort to explore this correlation further, we have calculated the average maximum value of the $z$ component, $< z_{max} >$, of 100 chaotic orbits (trapped and escaping) for each value of the energy $h$. Figure 11a depicts a plot describing the evolution of the average value of the maximum $z$ component, $< z_{max} >$, for each value of the energy $h$. In this case, the average value $< z_{max} >$ corresponds to the trapped chaotic orbits. One may observe, that the average value $< z_{max} >$ decreases as the value of the energy increases. Similarly, in Figure 11b we can see the same plot but in this case corresponds to the fast escaping orbits. Once more, the average value $< z_{max} >$ decreases as the energy $h$ increases. Here, we must point out that in the case of the fast escaping orbits the decrease is much more rapid than in the case of the trapped orbits shown in Fig. 11a. From the plots of Fig. 11a-b we can justify and also explain the behavior of the escape period which was discussed in Fig. 8a-b. As the average value $< z_{max} >$ decreases in both cases (trapped and fast escaping orbits), this should explain the same decrease of the escape period shown in Fig. 8a-b respectively. In particular, the pattern of Fig. 11a-b is very similar to those in Fig. 8a-b. Moreover, we see that in the case of the fast escaping orbits (Fig. 11b) the $< z_{max} >$ has much more lower values than in the case of the trapped chaotic orbits (Fig. 11a). Especially in Fig. 11b the values of $< z_{max} >$ are 10 to 40 times smaller than those of Fig. 11a. This means that the fast escaping orbits of the dynamical system move very close to the galactic plane, with very low values of the $z$ component, almost parallel to the escape channel of the open ZVC. Consequently, these fast escaping orbits can find easily and much more quickly the exit gap and therefore, this explains their small values of the escape period. On the contrary, our numerical calculations suggest that the trapped chaotic orbits of the dynamical system, have much larger average values of $< z_{max} >$. Thus, these orbit must spend vast time intervals inside the ZVC (much larger than the age of the Universe), before they eventually find the exit and escape. Therefore, we consider these chaotic orbits as ``trapped chaotic orbits".
\begin{figure*}[!tH]
\centering
\resizebox{\hsize}{!}{\rotatebox{0}{\includegraphics*{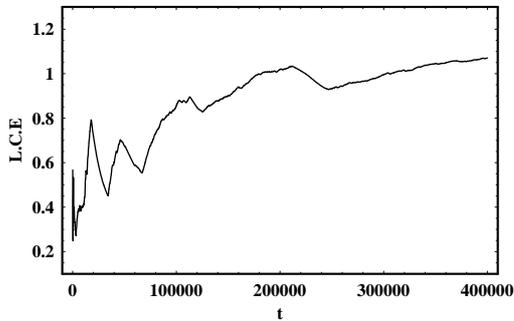}}\hspace{5cm}
                      \rotatebox{0}{\includegraphics*{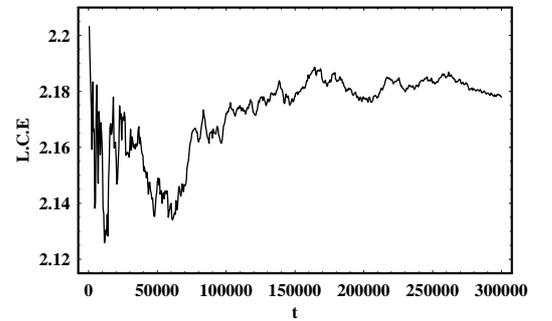}}}
\vskip 0.1cm
\caption{(a-b): Evolution of the the L.C.E for the trapped chaotic orbit producing the chaotic region in (a-left): Fig. 4c and (b-right) Fig. 4d.}
\end{figure*}
\begin{figure}[!tH]
\centering
\resizebox{\hsize}{!}{\rotatebox{0}{\includegraphics*{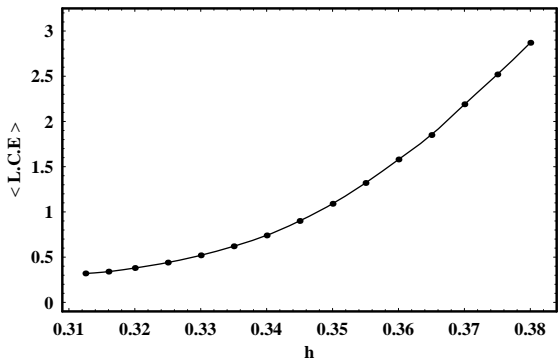}}}
\caption{A plot of the relation between the mean value of the L.C.E and the energy $h$.}
\end{figure}

It would be of particular interest to have a better estimation of the degree of chaos for the chaotic orbits, in each case. For this purpose, we have computed the Lyapunov Characteristic Exponent (L.C.E) (see Lichtenberg \& Lieberman 1992 for details). Figure 12a shows the L.C.E for the orbit producing the chaotic sea shown in Fig. 4c. The orbit stays inside the ZVC for about $4.2 \times 10^{12}$yr before escapes to infinity. The L.C.E of this orbit, is of order of unity. Thus, we are in the case of ``fast chaos" (see Caranicolas, 1990 and references therein). Figure 12b shows the time evolution of the L.C.E of the orbit producing the chaotic region shown in Fig. 4d. Here the orbit escapes after a time period of about $3.2 \times 10^{12}$yr. In this case, we have again fast chaos but the value of the L.C.E is much more larger. Figure 13 shows a plot of the relation between the average value of the L.C.E and the energy $h$. Here we must point out, that in general different chaotic orbits have different values of the L.C.E. As we have in all cases regular regions and only one unified chaotic domain in each $\left(r,p_r\right)$ phase plane of Fig. 4a-d, we calculate the average value of the L.C.E by taking 1000 orbits with different and random initial conditions $\left(r_0,p_{r0}\right)$ in the chaotic sea (see Caranicolas \& Zotos, 2011a). Note that, all calculated L.C.Es were different on the fifth decimal point in the same chaotic region. Therefore, the mean value of the L.C.E, is a representative value of the degree of chaos for each value of the energy $h$. Here we must point out, that the obtained average value of the L.C.E was calculated by choosing both trapped and fast escaping chaotic orbits. The exact nature of the escape of the chaotic orbit (fast or not) does not add any particular difference to the value of the L.C.E therefore, we do not need two different diagrams regarding separately trapped and fast escaping chaotic orbits, as we used in order to describe the escape period in Fig. 8a-b. In all cases, the values of the energy $h$ and the angular momentum $L_z$, are chosen as described in Section 2. We observe, that the $< L.C.E >$ increases drastically as the value of the energy increases. In all cases the ZVC is open, except in the case when $h=0.3125$, where the value of $L_z=0.01$ was taken. This was done, because for this value of the energy, for all values of $L_z$ the ZVC is closed.

If we combine the results from the phase planes shown in Fig. 4a-d together with the plot shown in Fig. 13, we can conclude that, as the values of the energy $h$ and the angular momentum $L_z$ increase the chaotic region in the phase plane decreases. At the same time however, as we can see from Fig. 13, the degree of chaoticity also increases drastically. Therefore, we can say that as the chaotic domain is reduced, this has as the result the increase of the degree of chaos. In other words, the more confined is a chaotic region in the phase plane, the more stronger is the degree of chaos in this region.

\section{Discussion and conclusions}

In the present article, we investigated the nature of the orbits of the stars, in a time independent galactic type potential. This potential can be considered to describe the motion in the meridian $(r,z)$ plane near the central parts of an axially symmetric galaxy. Our gravitational model is valid only when $r \leq 1.5$. The quartic terms in equation (1), become predominant as soon as we go beyond the above limit and therefore the model becomes unrealistic and unphysical. Using the conserved component of the angular momentum $L_z$, the three dimensional motion can be confined and studied in the meridian $(r,z)$ plane, which rotates differentially at an angular velocity given by equation (5).

We studied the phenomenon, where stars escape from the central parts of a galaxy. This phenomenon has been an active field of research over the last decades. Contopoulos (1990) using a simple model with two degrees of freedom and with energies larger than the escape energy, found that most orbits of the system escape fast but there are also orbits which escape only after an arbitrarily long time or do not escape at all. Similarly, Contopoulos and Kaufmann (1992) studied the nature of orbits in a Hamiltonian which has four channels of escape. They located the ``basins" of escape towards different directions in the cases of fast or slow escape, for various values of the perturbation parameter. Moreover, the work of Kandrup et al. (1999) summarizes a study of the problem of escapes of energetically unbound orbits in strongly non-integrable Hamiltonian systems of two degrees of freedom as an example of phase space transport in complex systems. This work has led to two significant conclusions: (i) When evolved into the future, ensembles of orbits of fixed energy often exhibit a rapid approach towards a constant escape probability $P_0$, the value of which is independent of the details of the ensemble and exhibits interesting scaling behavior. Moreover, the values of the critical exponents appear to be relatively insensitive to the choice of Hamiltonian and (ii) at later times, the escape probability decreases in a fashion which, for at least one model system, is well fit by a power law. This nontrivial time-dependence is attributed to the fact that the possibility of escape to infinity is controlled by the cantori, which can trap chaotic orbits near regular regions for extremely long times. Fukushige \& Heggie (2000) modeled a cluster as a smooth potential plus the steady tidal field of the Galaxy. In this model there is a minimum energy below which stars cannot escape. Above this energy, however, the time-scale on which a star escapes varies with the orbital parameters of the star (mainly its value of energy). This time-scale was quantified and estimated, with both theoretical arguments and computer simulations. Within the limitations of the model, it was shown that the time-scale is long enough to complicate the interpretation of full $N$-body simulations of clusters and that stars above the escape energy may remain bound to the cluster for about a Hubble time. A detailed study regarding the dynamics of the outer parts of barred galaxies beyond corotation was made by Contopoulos \& Patsis (2006). They found that in the outer regions of barred galaxies beyond corotation, there are three types of orbits: (1) ordered (periodic or quasi periodic), (2) chaotic and (3) escaping. Papadopoulos \& Caranicolas (2007) explored the character of orbits in a ``bare" Seyfert 1 dynamical model, when the external perturbation is strong enough in order to have open Zero Velocity Curves. In this model the majority of orbits escape to infinity but there are also orbits which are trapped and do not escape at all. Thus, it is evident that our results coincide with the outcomes of previous related work, pointing out that in a Hamiltonian system with escape channel, there are two kinds of orbits (i) escaping orbits and (ii) trapped orbits.

In the current research, we focused our study in the behavior of orbits when the Zero Velocity Curve (ZVC) of the dynamical system is open and therefore orbits can escape to infinity. This happens when $h > h_{esc}$, that is the case in which stars hold values of energy larger than the energy of escape. In this work we studied the possibility of escape in a dynamical system with two degrees of freedom and we have obtained quite striking results. Our results could be compared with outcomes from other studies about the non integrability of the $J_2$ problem (Irigoyen \& Sim\'{o} 1993) ar about the central manifolds and the destruction of the KAM tori in the planar Hill's problem (Sim\'{o} \& Stuchi 2000).

Using the fixed values of the parameters given in Section 2, we obtained the relation (7) which connects the angular momentum and the energy of escape. Our numerical calculations show, that all tested regular orbits do not escape at all from the system. When we refer to regular orbits, we mean both the basic orbits of the system and also orbits correspond to secondary resonances. Thus, we conclude that these orbits are trapped regular orbits. Moreover, the percentage of the regular orbits increases as the values of the energy $h$ and the angular momentum $L_z$ increase. More correctly, the whole area of regular orbits, in the $\left(r,p_r\right)$ phase planes increases as we proceed to higher values of the energy. At the same time, the area in the same phase planes corresponds to chaotic orbits decreases. Note that, in all cases studied, the ZVC is open and the values of the pairs $\left[h, L_z \right]$ were chosen, using the algorithm we described in Section 2.

Our numerical experiments, suggest that in addition to the regular orbits, for a particular pair of $\left[h, L_z \right]$, there are also two kinds of chaotic orbits. Chaotic orbits which spend large time intervals inside the ZVC before they escape and chaotic orbits which escape vary fast through the escape channel. It was observed, that about $62\%$ of the total tested chaotic orbits stay inside the ZVC for time intervals which are at least 100 times larger than the age of the Universe. Therefore these orbits can be considered as non escaping. On this basis, we are experienced a phenomenon of ``trapped chaos", that is when we have chaotic orbits which are trapped inside to an open ZVC. The rest $38\%$ of the tested chaotic orbits, correspond to orbits which have small escape periods and therefore can be regarded as fast escaping orbits. The main characteristic of these orbits, is that they have initial conditions very close to the critical boundary of the Lyapunov orbit. This is the boundary that orbits should be able to penetrate in order to escape to infinity. Here, we must point out that the fact that orbits with initial conditions very close to the Lyapunov orbit escape fast, is not a necessary condition for fast escape in general (e.g. in Fig. 9d we may consider the orbit starting close to the left boundary of the ZVC, far from the Lyapunov orbit and this orbit escapes fast). The numerical experiments indicate that the average value of the escape period of the chaotic orbits (trapped or fast escaping) strongly depends on the average value of the maximum $z$ component of the orbits. Our main conclusion, is that eventually all chaotic orbits, sooner or later, will escape from the system.

In order to have a better estimation of the degree of chaos for the chaotic orbits in each case, we computed the average value of the Lyapunov Characteristic Exponent. Our results indicate, that as the chaotic region is reduced, this has as a result the increase of the degree of chaos. In other words, the more confined is a chaotic region, the more stronger is the degree of chaos in this region. Especially in the case of trapped chaotic orbits, we deal with ``trapped fast chaos", as in most cases the value of the computed L.C.E is much more larger than the unity. When we state that the L.C.E has a value much more larger than the unity, we mean that the L.C.E is larger than $10^{-7}$ yr$^{-1}$, that is the inverse value of the time unit.

As the phenomenon of the trapped chaotic orbits is of particular interest, we shall stay in the subject with some more comments. In the case of trapped regular orbits, for values of energy larger that the energy of escape, one could say that these regular trapped orbits may have an additional third integral of motion, that keeps them inside the open ZVC forever. Of course, this can not mentioned in the case of trapped chaotic orbits, because chaotic orbits could not have any further integral of motion. So what remains, is a kind of ``stickiness" (see Karanis \& Caranicolas 2002), with a time of sticky period, larger than the age of the Universe!

We consider the outcomes of the present research, to be an initial effort, in order to explore the orbital structure of this interesting dynamical system in extensive detail in a future paper. As the results are positive, further investigation will be initiated in order to explore all the available phase space and also to cover the whole range of the values of the main involved parameters, such as the value of the energy $h$ and the value of the angular momentum $L_z$.

\section*{Acknowledgments}

The author would like to express his thanks to the anonymous referee for his careful reading of the manuscript and also for his very useful suggestions and comments, which improved greatly the quality and the clarity of the present paper.

\end{document}